\documentstyle[11pt,aaspp4]{article} 
\newcommand{\kmsM} {km~s$^{-1}$~Mpc$^{-1}$}
\newcommand{\subsun}{\mbox{$_{\odot}$}}
\newcommand{\etal}{{\it et al.\/}}

\lefthead{Cohen {\it et al.} }
\righthead{HDF Redshift Survey}

\begin{document}

\title{Caltech Faint Galaxy Redshift Survey XIII: 
Spectral Energy Distributions for Galaxies
in the Region of the Hubble Deep Field North\altaffilmark{1}}

\author{Judith G. Cohen\altaffilmark{2} }

\altaffiltext{1}{Based in large part on observations obtained at the
	W.M. Keck Observatory, which is operated jointly by the California 
	Institute of Technology, the University of California
        and the NASA,}
\altaffiltext{2}{Palomar Observatory, Mail Stop 105-24,
	California Institute of Technology, Pasadena, CA \, 91125}

\begin{abstract}

%
%
We introduce a new empirical function for modeling
the spectral energy distributions of galaxies.  We apply it to
a sample of 590 galaxies in the region of the HDF with $z < 1.5$ using
multi-color photometry
with wide wavelength coverage combined with spectroscopic redshifts
from our 93\% complete $R$-selected redshift survey there.  We find
the following:
\begin{itemize}
\item  As expected, galaxies with strong signs of
recent star formation (i.e. those which show emission lines) have 
bluer continua in both the rest frame UV and the optical/near-infrared.
\item The redder galaxies tend to be more luminous.  Although galaxies 
with strong absorption lines and no emission features
are $\sim$15\% of the total sample with $0.25 < z < 0.8$, they are 
$\sim$50\% of the 25 most luminous galaxies in the sample at rest-frame $R$.
\item  The SEDs of actively star forming galaxies
become bluer in the mean in the rest-frame UV at higher redshift,
which trend might arise from SED modeling errors.
Aside from this, we discern no change
with redshift in the relationship between SED characteristics
and galaxy spectral type based on the strength of narrow emission and 
absorption features.
\item Combining with similar work at higher
and lower redshift, the bluest galaxies 
have indistinguishable
spectral energy distributions in the rest frame ultraviolet
over the redshift regime 0 to 3.  
\item There is no evidence in our $R$-selected sample that supports
the existence of a substantial population of very dusty star forming
galaxies at $z \lesssim 1.5$.
\item  Our ability to predict the mid-IR flux using the UV/optical/near-IR
SEDs is limited.
\item The  potential accuracy of photometric
redshifts, bearing in mind that a break at 4000\AA\ must be detectable
to within the errors of the photometry to assign a photo-$z$ for
galaxies in this redshift regime, is evaluated.
\item The rest frame $K$-band luminosity as a function of redshift clearly 
shows a gradual change in the population
of various types of galaxies, with star forming galaxies becoming
both more luminous and a larger fraction of the total population 
at higher redshift.
\item The overall pattern of the $L(K) - z$ relationship
suggests that passive
evolution at constant stellar mass is a good approximation to the actual
behavior of at least the most luminous galaxies in
this large sample of galaxies in the region of the HDF
out to $z \sim 1.5$.  
\end{itemize}

\end{abstract}


\section{Introduction}

In the present paper, we combine the results of our redshift
survey in the region of the Hubble Deep Field North (henceforth
HDF) (Cohen \etal\ 2000) 
with multicolor photometric databases to derive the 
rest frame spectral energy distributions
(henceforth SEDs) of the sample's galaxies.  
After defining the form of the SED we adopt, and exploring the
limits of validity thereof, we concentrate on what can be determined from the
behavior of the SED parameters themselves as a function of
redshift, galaxy spectral type (i.e. the presence or absence
of key emission and absorption features) and luminosity.
A comparison of the observed behavior of the SED parameters
with the predictions of
galaxy evolutionary synthesis models reveals important differences. 

The SEDs of the star forming galaxies
are then used to constrain possible presence of
a substantial population of dusty starburst galaxies and to test our
ability to predict the mid-IR thermal emission from dust in
star forming galaxies in \S\ref{ISO}.  
Constraints on the variation in internal reddening
from galaxy to galaxy of a given spectral type are derived as well.
We compare the rest frame UV SEDs of the bluest star forming galaxies
over the regime $z = 0$ to $z \sim 3$.

We also evaluate the ability of
photometric redshift schemes to discern the 4000\AA\ break given realistic
SED distributions and errors characteristic of ground based photometry
in \S\ref{photoz}.
We explore in \S\ref{klum} the conversion from luminosity in the rest-frame
infrared into total stellar mass, which relies
on a calibration from models of the integrated light of evolving galaxies.
A short summary concludes the paper.

As in earlier papers in this series, we 
adopt the cosmology H$_0 = 60$ \kmsM, $\Omega_M = 0.3$, 
${\Omega}_{\Lambda} = 0$.
Over the redshift interval of most interest, a flat universe with
$\Omega_{\Lambda} = 0.7$ and a Hubble constant of H$_0 = 67$ \kmsM\
gives galaxy luminosities very close to those derived below.

\section{The Sample of Galaxies \label{sample} }

We have recently completed an extensive redshift survey in the region of the
HDF.  The survey is magnitude-limited, with a selection
at $R$, and objects are observed irrespective of morphology.
Redshifts have been obtained for more than 92\% of the objects in the
HDF (Williams \etal\ 1996) with $R < 24$, and for more than 92\% of the objects
within a circle whose diameter is 8 arcmin centered on the HDF with $R < 23$.
The redshift catalog, presented and described in detail in
Cohen \etal\ (2000), contains 671 entries.  
Hogg \etal\ (2000) (henceforth H00) 
present a four filter photometric catalog for the region of the HDF
including the Flanking Fields, 
with images in $U_n$, $G$ and $R$ contributed by C. Steidel,
and it is their $R$ catalog
that was used to define the samples for the redshift survey.

Appendix A gives an update to this survey with a total of 
58 new redshifts, including 5 in the HDF itself.  With this
addition, the completion of the redshift sample
in the HDF itself to $R \le 24$ is now 95\%, while the completion
in the Flanking Fields to $R < 23$ is now 93\%.

Here we use only the sample of galaxies with $0 < z < 1.5$,
eliminating the Galactic stars, 
the two broad-lined AGN with $z < 1.5$, and the higher redshift ($z \sim 3$)
objects 
\footnote{The two AGNs with $z < 1.5$ are sometimes included in the
figures.}
leaving 590 galaxies.  We do this as  our ground based photometry is
measuring largely the rest frame UV for $z \sim 3$ galaxies and
does not provide enough coverage in the
infrared to yield an adequate determination of rest frame optical SEDs.

The rest frame UV
is very poorly determined in the nearest galaxies with $z < 0.25$
when photometry is limited to
ground based photometry without access to space based
mid and far UV observations.
Hence they are eliminated in the rest of this paper. 
Although SEDs are determined and tabulated for galaxies with $z < 0.25$
here, we will utilize 
in the subsequent discussion the sample of galaxies restricted
to the range $0.25 < z < 1.5$,  which contains 552 galaxies of
which 107 are in the HDF itself.

%
%

The results presented throughout this paper are robust to the elimination
of the fainter galaxies in the HDF, where the sample is deepest.

\subsection{The Redshift Ranges}

The present sample of galaxies in the region of the
HDF with $0.25 < z < 1.5$ and with SEDs considered reliable
contains 519 galaxies,
of which 105 are in the HDF itself.
\footnote{In \S\ref{problems} we find that 33 of the galaxies
have SEDs not considered reliable.}
These are divided into four redshift ranges:
``low'' ($0.25 \le z < 0.5$), ``mid'' ($0.5 \le z < 0.8$),
``high'' ($0.8 \le z < 1.05$), and ``highest''
($1.05 \le z \le 1.5$).  In the ``highest'' redshift range, the
assignment of galaxy spectral classes is less accurate
(see the discussion in Cohen \etal\ 1999a) and the redshift completeness
is affected by the fact that 
the intrinsically strong [OII] 3727\AA\ emission
line is shifted into the region  beyond 7500\AA, where strong night
sky emission lines make detection of faint emission 
lines from distant galaxies more difficult.

\section{The Derivation of the SEDs}

The derivation of the spectral energy distribution for a galaxy 
requires a set of 
photometric measurements for the object
with $N$ filters, a set of effective wavelengths
and flux zero point calibrations for the filters, and a model for
fitting the SED.  Our goal is an overall characterization of the SED
suitable for the determination of luminosity functions given the
fact that we are dealing to a large extent with ground based photometry
of faint galaxies.

We are 
trying to obtain a set of redshift independent parameters characterizing
the rest frame SED of a galaxy from a set of observations made at filter
bandpasses fixed in the observed frame.  We would like to be able to 
use our SED model parameters 
to predict fluxes of galaxies from 2400 \AA\ to 2.2$\mu$ in the
rest frame over the redshift range
$0.25 \le z \le 1.5$ with an accuracy of 25\% (0.1 dex). 

The complexity of the SED model to be used here is restricted
by the limited photometry available for these objects,
both in terms of accuracy for these faint galaxies
and also in wavelength sampling.  Were we dealing with 
HST data only (for the same set of filters) or
with much brighter galaxies,
a much more sophisticated approach such as that of Budav\'ari \etal\ (2000)
would be justified.

\subsection{The Sources of Photometry}

Since the catalog of H00 was used to define the sample for the redshift
survey of Cohen \etal\ (2000) and since it provides four filter photometry
($U_n$, $G$, $R$ and $K_s$)
for the entire sample, we adopt it as the primary photometric source.
The H00 photometry is consistent to within the uncertainties with the more accurate
photometry of Williams \etal\ (1996) for objects in the HDF itself for
the optical colors.
In the Flanking Fields around the HDF, we
supplement the photometric database of H00 with 
the $I$ and $K$ measurements of Barger \etal\ (1998).
We utilized primarily 
$I$ from Barger \etal\ (1998)
to fill in gaps in the wavelength coverage of H00.  For
the faintest objects (in the infrared) in the Flanking Fields, we used $K$
as well, as the limiting magnitude of the Barger \etal\ (1998) database
is fainter at $K$ than that of H00.   The $U, B, V, R$ bands of
of Barger \etal\ are used to supply added confidence
in the $U_n,G,R$ photometry of H00.  Thus, for an object in the
Flanking Fields, there is a maximum of 10 observations, with two independent
measurements for three colors ($U, R$ and $K$), hence seven distinct
filter bandpasses.
  
Within the HDF, in addition to the photometric
catalog of H00, we also use the $I,J,H,K$ photometry of
Fernandez-Soto, Lanzetta \& Yahil (1999).
Their catalog is based on 
the HST images of the HDF with the F814W filter for $I$, and on 
their analysis of IR images of the HDF from KPNO 
described in Dickinson (2000).
The red end of the SED is thus better determined for galaxies in the HDF
itself, due to the existence of the $J$ and $H$ photometry and the very
high quality of the F814 HST images compared to ground based $I$ images.

The H00 and Fernandez-Soto, Lanzetta \& Yahil (1999) 
photometric databases appear to be consistent with each other.
They both use the Sextractor code (Bertin \& Arnouts 1996) for
object selection, and both have a scheme to handle extended objects,
yet still preserve accuracy for the smallest and faintest galaxies.
The unpublished photometry of Barger \etal\ (1998)  
treats the extrapolation to the total magnitude of extended objects
differently from the procedure adopted in H00. 

We have
adopted a single effective wavelength for each filter for 
all objects irrespective of their spectral indices, ignoring
any dependence of $\lambda_{eff}$ on the color of the object.
This is a reasonable
assumption given the precision 
of our photometry for faint objects and the desired
precision of the SED indices to be derived.
The effective wavelengths of the various filters are given in 
an appendix.

Once the effective wavelengths were defined, the zero points
of the observed flux $f_{\nu}$
for  the Vega-relative photometry were taken from 
Fukugita,  Shimasaku \& Ichikawa (1995).
The HDF photometry of Fernandez-Soto, Lanzetta \& Yahil (1999) is
already presented in the form of observed flux per unit frequency $f_{\nu}$.

A small extinction  of 
E($B$-$V$) = 0.012 mag was adopted based on the maps of 
Schlegel, Finkbeiner \& Davis (1998); appropriate corrections were
applied to each filter.

Our photometric database covers $U$ through $K$ in the observed frame.
The red end of this corresponds to 0.9$\mu$ in the rest frame of
the highest redshift considered here and to $\sim1.3\mu$  
for a more typical $z\sim0.7$ galaxy.  Prediction of fluxes
in the rest frame redward of 0.9$\mu$ requires trusting the parametric
form of our adopted SED model described below
for the highest redshift galaxies in the sample.

\subsection{The Dual Power Law Model for the SEDs \label{2p_sed_model} }

We initially adopted the same form for SEDs
as in Cohen \etal\ (1999a),
namely we assume that the emitted luminosity per unit frequency
in the rest frame over the wavelength regime 0.2 to 1.6$\mu$ can be
represented by a power law whose index may change at
4000\AA.  Thus $L_{\nu}$, with units of Watts/Hz, is assumed
$\sim \nu^{-\alpha}$ with an index in the region
redward of 4000\AA\ (in the rest frame) denoted $\alpha_{IR}$ and an index
in the rest frame UV of $\alpha_{UV}$.  Three parameters are required
to characterize each SED.  We refer to this as the 2p SED model.

Several tests of the validity of this model have been made
to establish the wavelength range over which it can represent
galaxy SEDs with the requisite degree of precision.  Both 
SEDs from calculations of the integrated light of evolving galaxies
and SEDs
interpolated from spectrophotometric and broad-band photometric
observations of nearby galaxies have been used in these trials.
In the first test, we fit the galaxy evolution models of Poggianti (1997),
using the filter transmission curves for the set of filters for
which photometric catalogs exist in the region of the HDF.
Figure 1a illustrates SEDs from Poggianti of
a local elliptical, Sa and Sc galaxy (henceforth denoted
as the standard set of test SEDs) observed at $z=0.6$.
The observed filter bands are the large circles plotted
at their appropriate rest wavelengths.  
While the spectral
region beyond the 4000\AA\ break for all the models from Poggianti
that were examined is well represented by a single power law with
deviations never exceeding 0.1 dex,
the 2p model systematically predicts too much flux beyond 1.7$\mu$.

Tests of the 2p model were also carried out
with galaxy SEDs predicted from Worthey (1984) and with
a set of SEDs of nearby galaxies of various morphological types
constructed by G. Neugebauer (private communication, 2000) based on
observations in the the UV
and optical assembled by Coleman, Wu \& Weedman (1980) supplemented 
with broad-band
IR photometry from Aaronson (1978) and from Frogel \etal\ (1978).  In
all cases the 2p SED overpredicts the flux in the near infrared.

We compare the predicted flux at an observed wavelength 
of 3.2$\mu$ from our SED model with those
galaxies in the HDF with detections at that wavelength from
Hogg \etal\ (2000). Given the typical redshift of these
galaxies of $z\sim0.5$, these observations have a typical
rest wavelength of 2.1$\mu$. Again, the 2p model for the SED
overpredicts the observed flux for the 9 galaxies with detections
by about a factor of two.

This deviation in the near infrared between the power law and the
galaxy flux presumably largely arises because  
one is then well into the long wavelength tail of a black-body 
distribution, where even for $T$ as low as 4000K, $f_{\lambda}$ is no 
longer rising rapidly with $\lambda$.  

In the UV, the high spectral resolution of the
Poggianti model grid demonstrates the existence of a different posible concern.
Many of Poggianti's models show  f$_{\lambda}$ 
falling shortward of
of the 4000\AA\ break, and then rising again as flux from the youngest
and hottest stars begins to contribute substantially.  This curvature
obviously cannot be reproduced well by a single power law flux
in the UV.  

These tests demonstrate that the 2p model SED is at best
marginally adequate for our purposes.

\subsection{A New Empirical Model for Galaxy SEDs \label{sBBfits} }

To avoid the difficulties of 2p SED model described above, 
we have
developed a new model for galaxy SEDs.  One natural way to introduce the
desired curvature in the optical and near IR spectra region is
to replace the power law with a black
body function.  However, the peak of the Planck function itself is too narrow 
to give a good fit to actual galaxy SEDs, as one might expect for
a composite stellar population containing stars with a range
of effective temperatures.
The empirically determined function we adopt as our SED model, 
which provides a simple but versatile parameterization of
galaxy SEDs with parameters that are physically motivated, 
has a mathematical form which we denote as 
a ``stretched black body'' (henceforth sBB).  We apply
the sBB function to fit the optical/near IR,
while we retain a power law fit to the UV.  

A wavelength 
$\lambda_m$ is
specified as the wavelength at which the fit function changes.
The ``stretched wavelength'' $\lambda_s$ corresponding to a
rest frame wavelength $\lambda_0$ is then defined as 
$$\lambda_s = \lambda_m + (\lambda_0 - \lambda_m)/f, $$
for $\lambda_0 \ge \lambda_m$, where $f$ is the stretch factor.
To evaluate the sBB Planck function, the
``stretched wavelength'' is converted to a frequency in the usual
way, $\nu_s = c/\lambda_s$, and then used in the calculation.
This applies to both the $\nu^3$ and exponential part of the Planck
function, as well as the $\nu$ multiplier required to obtain luminosities.
The temperature used in the Planck function is denoted as $T(sBB)$.
The details of the fitting procedure are given in an appendix.

It is important to note that in this model, unlike in the
2p SED model, the
flux at $\lambda_m$ from the
red fit and from the blue fit is not automatically constrained to be 
the same.  This offers the possibility of direct measurement of
the the 4000\AA\ break which in cool stars is due to enhanced
absorption by metal lines and in hotter stars by the Balmer jump.
An sBB fit thus has four parameters, $\alpha_{UV}$, $T(sBB)$, 
the blue side luminosity and the red side luminosity, both at $\lambda_m$.
As described in the appendix, depending on the number and distribution
with wavelength of the available filter bandpasses (i.e. photometric catalogs),
the two luminosity parameters are not 
always independent.

There are two constants in a sBB fit,
$\lambda_m$ and the wavelength stretching factor $f$.
Their values were determined by optimizing $\chi^2$ 
when a sBB fit was applied to the standard test set of
Poggianti's (1997) local galaxy SEDs
observed at $z=0$ and at $z=0.6$.  The resulting
choices adopted henceforth for sBB fits are
$\lambda_m = 4050$ \AA\ and $f = 1.6$.

Figure 1b shows the sBB fit to the set of standard
test SEDS,  i.e. the same SEDs as is shown in Figure 1a. 
The sBB fit solves
the problem of overpredicting the rest frame
flux at $K$ and, with only one extra parameter, the 
sBB model shows a much better fit overall to 
predicted galaxy SEDs, as well as allowing a measurement of
the 4000 \AA\ break.

\subsection{The Stability of the SED Parameters with Redshift
 \label{sed_model_tests}}

The model SED must 
measure the same values for its parameters 
irrespective of the redshift of a galaxy.
Otherwise the flux cannot be predicted reliably at a fixed
rest wavelength.  This  is
particuarly difficult at $K$, as we are then
extrapolating beyond the rest frame wavelength
range of the available photometry for $z > 0$.
As pointed out by the referee, the 2p SED model
fails this test.  Figure 2 shows the output
parameters from this model, $\alpha_{UV}$ and
$\alpha_{IR}$, applied to the set of standard local
test SEDs as the redshift of observation
is varied from 0 to 1.5.  (The dependence of $\alpha_{UV}$
in the region
$z < 0.25$ should be ignored; see \S\ref{sample}.)
Note the large changes in output parameters
for the 2p model SED as the redshift at which the
galaxy is observed is varied.

This translates into predictions of flux using
the 2p SED model which are,
at the extremes of the rest wavelength range covered,
so different from the those of the actual SED that our
accuracy tolerance is grossly exceeded.
Figure 3 shows the deviation from
the fluxes of the standard test set of 
local SEDs 
predicted by the 2p 
SED model as the observations are carried out
over the redshift range of interest,
from $z=0$ to $z=1.5$.  Results at four rest wavelengths,
the two wavelength extremes of our range,
2400 \AA\ and 2.2$\mu$,
as well as at 
two intermediate wavelengths, 4350 and 7900 \AA,
are illustrated.  As expected,
the ability of the 2p SED model to predict the flux at the
extreme ends of the wavelength range is poor, although
in the central part of the wavelength range it is reasonably good.

Figures 4 and 5 show the equivalent tests applied
to the sBB model.  Note that  
Figures 3 and 5 are directly comparable; the
same rest wavelengths, axis scales and plot symbols are used in both cases.
We see that the predictive ability of the sBB fit
is quite good; over the full redshift range
$0 < z < 1.5$, the rest frame flux at $K$ is
predicted to with $\pm0.1$ dex (our required
tolerance), while the rest frame flux at 2400\AA\
is predicted almost as accurately for $0.25 < z < 1.5$.

At this point, with the sBB fits,
the worst remaining problem is now in the rest frame UV (see figure~4).
The overall trend towards measuring a bluer $\alpha(UV)$ as $z$
increases, particularly for the Sa and Sc SEDs, is probably due
to an attempt to fit a curved SED which is ``convex upward''
as the contribution from younger stars increases and changes the
curvature.  In spite of this, we note that the flux prediction of the sBB model
at 2400\AA\ shown in Figure~5
is still close to or within our tolerances.

These tests have been repeated for the sBB model omitting the $J$ and $H$ 
band coverage, as occurs in the Flanking Fields of the HDF
in contrast with the HDF itself, where $J$ and $H$ photometry
is available.  The sBB fits are still
very good and remain within the specified tolerance until
$z > 1.1$.

\subsection{Final Comments on the SED Models}

Extrapolation beyond rest-frame $K$ or blueward of 2400 \AA\ 
of our adopted form for galaxy SEDs is not appropriate.

Additional data on the 1.5 to 5$\mu$ SEDs of
nearby galaxies at better spectral resolution
than is provided by standard broad band infrared photometry
would be helpful for improving the accuracy of galaxy synthesis models 
in the near IR.

\subsection{Reddening}

We have tried adding additional reddening to the SEDs of local
galaxies, both synthesized from models and observed.  
As one might expect, an additional reddening of $A_V = 1$ mag with the standard
Galactic extinction curve produces an increase in $\alpha_{UV}$ of $\sim$2
and an increase in $\alpha_{IR}$ of $\sim$0.5.  The fit over the wavelength
range of interest is degraded, although not
beyond the limit of acceptability until the additional $A_V$ reaches 2 mag.  
As we will see later,
the range of the indices $\alpha_{UV}$ and $\alpha_{IR}$
within each galaxy spectral type is sufficiently small that
a range in reddening within each galaxy spectral type
significantly exceeding $\sigma(A_V) = 1$ mag can be ruled out.
However, a much larger range in reddening can be tolerated if
it is grayer than the standard Galactic extinction curve,
and has, for example, the form advocated by Calzetti (1997).

\subsection{Construction of the SEDs}

The SEDs for the galaxies in the region of the HDF
were constructed by calculating for each filter bandpass 
for which photometry exists the quantity ${\nu}L_{\nu} = 4 \pi D_L^2 \nu f_{\nu} =
{\nu_0}L_{\nu_0}$, where $\nu$ is the observed frequency and
$\nu_0$ is the rest frame frequency. ($D_L$ is the luminosity
distance in our adopted cosmology.)  We
then shift the observed effective
wavelengths into the rest frame.  For each galaxy a plot was made of
the raw SED which was then inspected manually.  For about 10\% of
the objects, one deviant point, presumably corresponding to one bad
measurement, was adjusted.  In addition, the difference in large
aperture extrapolation between the H00 and Barger \etal\ datasets
had to be removed.  Our magnitude zero point is based on that of H00.

A fit to the set of $N$ values {${\nu}L_{\nu}$ for each galaxy 
was then attempted.   Details of fitting procedures and how the
luminosity at the matching point is handled are given in
an appendix.

The resulting parameters for each galaxy for the 2p SEDs
are the rest frame emitted luminosity
at $B$, $L(B) \equiv \nu L_{\nu}$ evaluated at $B$ in the rest frame
and the two power law indices. The range of $\pm1\sigma$ values
for $\alpha_{UV}$ was from 0.2 to 0.9, with a typical value $\sim0.5$
in the Flanking Fields and $\sim$ 0.4 in the HDF itself. 
For the sBB fits, the resulting parameters are $\alpha_{UV}$, $T$, 
and $L(\lambda_m:{\rm{blue}})$ and $L(\lambda_m:{\rm{red}})$.
These values are given in Table~1 for our sample.

\subsection{Special Cases \label{problems} }

A number of special cases arose for very faint or very crowded
objects in the Flanking
Fields.  If there was no detection of a galaxy
at 2.2$\mu$, then the median value of the IR parameter
for that redshift range and galaxy spectral type (see \S\ref{SEDs})
was adopted for the object.  A check was made to be sure that this
value was consistent with the limit of $K \sim 20$ of the H00
photometric survey and of the somewhat fainter limit for
the Barger \etal\ (1998) survey.

If the object was very crowded, usually the fainter of
a close pair on the sky, the H00 database only contains a $R$ mag,
which was estimated after the main database was assembled
during the cross checking of the redshift catalog
with the photometric catalog (see Cohen \etal\ 2000 for details).
In such cases, the relevant median was adopted
for each of the SED parameters.  The rest frame
$B$ luminosity was then the only parameter
calculated from the limited observations available.

There are a total of 33 such objects.  These galaxies are not included in Table~1 
nor in any plots of the spectral indices or
calculations of their properties, but
will be included in calculating the luminosity function in a future paper.

With two exceptions, we were able to determine reliable SEDs 
for all the galaxies with redshifts within the HDF itself.


\section{Properties of the SEDs\label{SEDs}}

We discuss the properties of the SEDs as a 
function of redshift,
luminosity, galaxy star formation rate, etc. 
To demonstrate that the results are robust,
we retain both the 2p and the sBB fits in the tables, but
since the sBB model for galaxy SEDs is clearly superior, only those fits are shown
in the remaining figures. We use the
galaxy spectral classification scheme defined in Cohen \etal\ (1999b), which
basically characterizes the strength of the strongest emission lines,
particularly [OII] at 3727 \AA, [OIII] at 5007 \AA\ and H$\alpha$ 
relative to the strong absorption features,
H and K of CaII and the normal absorption
in the Balmer lines.  To review briefly,
``${\cal E}$'' galaxies have spectra dominated by emission lines,
``${\cal A}$'' galaxies have spectra dominated by absorption lines,
while  ``${\cal I}$'' galaxies are of intermediate type.   Galaxies
with broad emission lines are denoted as spectral class ``${\cal Q}$''.
Starburst galaxies showing the higher Balmer lines 
(H$\gamma$, H$\delta$, etc.) in emission 
are denoted by ``${\cal B}$'', but for such faint
objects, it was not always possible to distinguish them from 
``${\cal E}$'' galaxies.
These classifications were assigned for the galaxies our sample in
the region of the HDF in Paper X (Cohen \etal\ 2000).  

To illustrate our SEDs applied to real galaxies,
Figure~6 shows the SEDs  for the five sources in the HDF detected by
Chandra (Hornschemeier \etal\ 2000) included in our sample.  The mid-IR
luminosities determined from ISO observations by
Aussel \etal\ (1999)
and the VLA radio luminosities inferred from the work of Richards \etal\ (1998) and Richards (2000)
are shown as well when the galaxies were detected. 
The  X-ray detections  
are in galaxies with spectral classes $\cal  Q$ (a broad lined AGN), 
$\cal  A$, $\cal  I$, $\cal  EI$
and $\cal  E$, i.e. three of the five do not show have strong narrow
emission lines, hence do not show evidence for 
a high current
rate of star formation.

\subsection{The Correlation between SED Parameters and Galaxy Spectral Types}

Table~2 gives the medians of $\alpha_{UV}$ and $\alpha_{IR}$
from the 2p model SED
over the four redshift ranges and for various 
spectral classes of galaxies.  Table 3 provides the same for
the parameters $\alpha_{UV}$ and $T$ from the sBB model. (log($T$) is
actually used.)
Only galaxies with reliable
redshifts (redshift quality class 1, 2, 4 or 6, see Cohen \etal\ 1999b
for definitions) are used, as is true for all subsequent
tables and figures.
\footnote{Unless otherwise specified, starbursts (galaxy spectral
type $\cal  B$) are included with the $\cal  E$ galaxies. Broad-lined
AGNs are excluded throughout.}
While the distribution of these indices is non-Gaussian, we have
used the first and last quartiles of the distribution in each case
to produce a corresponding $\sigma$ assuming a Gaussian distribution
prevails.

The dispersion for each galaxy spectral type is a combination of 
the intrinsic dispersion and
that induced by variations in internal reddening from galaxy to galaxy
as well as the known uncertainties in determining the SED parameters.
Making the absurd assumption that the first and last factors are negligible,
these values of $\sigma$ can be used to set firm upper limits
on internal reddening variations within galaxies.  
If one assumes a Galactic extinction law, then
$\sigma(A_V) \sim 1$ mag will reproduce the observed range
in $\alpha_{IR}$ for each of the three galaxy spectral classes.
Again assuming a Galactic extinction law, the dispersion in
reddening required to reproduce the range of $\alpha_{UV}$ is
about 1/3 as large.  The adoption of a grayer reddening curve
will allow a larger range in $A_V$. 

As expected, the galaxy
spectral classes are correlated with the overall SED shape.
``Bluer'' galaxies (those with bluer continua) tend to 
show have stronger emission lines,
while redder galaxies tend not to have detectable emission lines.
This is apparent in the local Universe, and we reaffirm this
again at much higher redshift.  This is true not only for
the rest-frame UV continuum, but also for the rest-frame Paschen continuum,
although the effect is smaller there.
The observation that galaxies with strong emission lines (i.e. those
with strong current star formation) are bluer is one found
by many other surveys, both locally and within the redshift range
under discussion here, e.g. Lin \etal\ (1996) for the
Las Campanas Redshift Survey, Ellis \etal\ (1996)
for the LDSS survey, Cowie \etal\ (1996), Hammer \etal\ (1997) for the CFRS.

Figure~7 shows a plot of the two sBB SED parameters $\alpha_{UV}$ and $T$ 
%
%
for galaxies in the ``mid'' and the ``high''
redshift ranges. 
This figure (when constructed using the 2p SED model parameters)
looks very similar to the corresponding figure
(Figure~5) of our analysis of data from our
survey field at J0053+1234 given in
Cohen \etal\ (1999a). As would be expected from Tables~2 and 3, galaxies whose
spectra are dominated by emission lines occupy a different area in this
plot than do the galaxies without detectable emission lines.
Again as expected, galaxies with strong emission lines are 
significantly bluer in the rest-frame UV and somewhat bluer in 
the rest frame optical/near IR, 
while galaxies with no sign of ongoing star formation are redder
in both in the rest frame UV and in the
Paschen continuum.

To reinforce this point,
a histogram of SED indices with galaxy spectral type 
for galaxies in the ``mid'' and ``high'' redshift range is
shown in Figure~8a, 8b.

\subsection{The Correlation between SED Parameters and Redshift }

Table~2 and Table~3 demonstrate
a decrease of $\alpha_{UV}$ with increasing redshift for $\cal  E$ galaxies
which is in the expected sense, i.e. higher redshift star forming
galaxies appear bluer in the rest frame UV than their local 
counterparts, probably due
to a higher contribution to the total galaxy luminosity from young stars, i.e.
a higher mean star formation rates per unit galaxy luminosity.
One may also attribute this change to the limited ability
of even the sBB model to produce for a fixed galaxy SED a constant value of $\alpha(UV)$ as
a function of redshift (see figure 4).

However, with the exception noted above, 
the UV and optical/near IR SED
indices ($\alpha_{UV}$ and $\alpha_{IR}$ or $T$)
appear to be constant with redshift
for each of the three major galaxy spectral classes,
${\cal A, I}$, and 
$\cal  E$, to within the uncertainties defined by assuming
that the scatter has a Gaussian distribution for the ``low'', ``mid''
and ``high'' spectral ranges.  The CFRS group (Hammer \etal\ 1997) also found only
modest evolution in mean colors
with redshift.
 
The rest frame optical/near IR continua of $\cal  E$ galaxies appear to become
somewhat redder at high redshift, but this is dependent on the median of
$T$ or $\alpha_{IR}$ for the
highest redshift bin.
It is not surprising that the $\cal  E$
galaxies in the 
the ``highest'' redshift range appear to have somewhat
redder rest-frame optical continua, as at that point
the galaxy spectral classification becomes unreliable, and the
$\cal  E$ class is a catch-all for a wide range of galaxies.  Similar concerns
manifest themselves for the highest $z$-range in which $I$ galaxies could
be detected.

This result is equivalent to stating
that the association between
the spectral indices that characterize the continuum slopes
of the median rest frame SED of galaxies and the presence of certain 
discrete narrow spectral features
(i.e. those that define the galaxy spectral classes ${\cal A, I}$, and 
$\cal  E$) is roughly invariant out to $z \sim 1.1$, with the
exception that actively star forming galaxies become bluer in the UV
at higher redshift.  

We can compare the behavior of the SED parameters with redshift
with that predicted by Poggianti's (1997) galaxy evolutionary synthesis
models.  We apply the same procedure to determine the power law indices
to these galaxy models, assuming that her E, Sa, and Sc models
correspond roughly to our galaxy spectral classes ${\cal A, I, E}$.
Her elliptical models show SEDs that change little with age,
at least to $z \sim 1$, in accordance with the data shown in Table~2.   
This corresponds
to the well known difficulty of determining the age of an old
stellar population from broad band photometry alone.

However, Poggianti's Sa and Sc models become bluer at all wavelengths within
the range of interest more rapidly between $z \sim 0$ and $z \sim 1$ 
than do the galaxy SEDs in our sample.  This difference is significantly
larger than the uncertainties.  

Galaxy evolutionary synthesis calculations involve many theoretical
inputs and many assumptions, with only a limited number of
constraints applying, mostly at $z=0$.  One expects that the
errors in stellar evolutionary tracks or in
our cosmology (i.e.
in the relationship between redshift and age) will not
be substantial.  
However, errors in the synthesis models arising from the  
forms adopted for the
star formation rate as a function of time (i.e. at $z > 0.5$) 
for the various current galaxy spectral classes 
may be more serious.

An interesting possibility for a partial
explanation is a scenario where as 
galaxies evolve in time, their SEDs age, and their 
morphological classifications may also evolve.  
Van den Bergh, Cohen, Hogg \& Blandford (2000), among others, present evidence
from an analysis of the morphology of this sample of galaxies 
on the HDF images taken by HST supporting
morphological evolution of galaxies with redshift.  

The most likely explanation of this difference between the behavior
with redshift of the SED parameters of our sample, divided in spectral 
classes, and predictions of galaxy evolution models is
the star formation rate of a particular model galaxy
from Poggianti (1997) varies so much
with time out to $z \sim 1.5$ that as their SEDs age  
the galaxies will shift between
the relatively small number of galaxy spectral classes used here.
This is an issue that will be explored in future work.

\subsection{The Correlation of Luminosity with Color}

In Cohen \etal\ (1999a) we found a tentative relationship between spectral
slopes and galaxy luminosity such that more luminous galaxies
are redder, suggestive of a continued correspondence at high $z$ 
to the well known
galaxy luminosity - mean metallicity relationship shown by early type
galaxies in the Local Universe.  (See,  for example,
the discussion of spiral galaxies in Zaritsky, Kennicutt \& Huchra 1994,
and for the dwarf galaxies in the Local Group see 
C\^ot\'e, Oke \& Cohen 1999).  The
large range in star formation rates among galaxies will also
contribute to a relationship between SFR and luminosity if most
star formation is now occuring within low luminosity galaxies.
Such galaxies would then appear bluer than high luminosity
more quiescent galaxies (see, e.g. Boselli \etal\ 2000).

We now examine our
much larger data set in the HDF to see if this trend persists.
We construct the rest frame optical luminosity using the parameters from the SED
fit for each galaxy.  For the 2p SED model, we use rest frame $R$,
while for the sBB model, we use $L(\lambda_m:{\rm{red}})$.
Even for the highest redshift
galaxies in this sample, rest frame $R$ lies within the range of
the existing broad band photometry, so no extrapolation of the SED
is required.

The results are given in Table 4 for the 2p SED model and
Table 5 for the sBB SED model. 
The luminosity characteristic of each redshift range
rises with increasing $z$ for the redshift range, as is expected since
ours is a magnitude limited sample.
There is a clear trend of galaxies with redder
UV power law indices (larger $\alpha_{UV}$) being more
luminous in every redshift range except in the highest range,
see Figure~9.
The increase in median rest frame optical
luminosity from the bluest to the reddest galaxies
in each redshift range is about a factor of 5.

The same trend is present for the IR power law indices, but the
increase is by a smaller factor, about 3, from the bluest to
the reddest galaxies.

The highest redshift range behaves anomalously, but that is not
surprising, as only the most luminous galaxies can be detected there
at such high redshifts, and the depth in luminosity of the sample
is small.  Furthermore, it is very difficult to assign redshifts
beyond $z \sim 1$ to systems dominated by strong absorption lines,
which are the most luminous galaxies in the lower redshift ranges;
any such galaxies may be so red as to 
fall below the $R$ cutoff of our sample.

To summarize, the reddest galaxies in both the rest frame UV and
optical/near-IR
tend to be the most luminous, at least to $z = 1.05$, after which, 
such a trend, if present, would be very hard to establish from our sample.
For example, of the 25 most luminous (in rest frame $R$) galaxies
with $0.25 < z < 0.8$, 12 have been assigned to galaxy spectral class
$\cal A$, but less than 15\% of the total sample within that redshift
range is assigned to this spectral class.

This is a well established trend in the local Universe, shown clearly
in the recent large samples from
the SDSS and 2dF surveys analyzed by Blanton \etal\ (2000) and by 
Folkes \etal\ (1999) respectively.

Since this is true in the rest frame optical, the effect will be even larger
at rest frame $K$, which is a better measure of the true luminosity
of the stellar system, unperturbed by contributions from a small number
of bright young (blue) stars.

We thus see a picture of quiescent non-star forming galaxies being
the most luminous at each $z$, while star-forming galaxies 
are bluer and also in each of the redshift ranges up to $z \sim 1.1$
have on average significantly lower
rest frame optical (and even more so $K$) luminosities.

\section{Properties of the SEDS of the $\cal  E$ Galaxies}

Our selection of star forming galaxies is complete, in that
all objects with strong emission lines are included in the $\cal  E$
spectral class. 
We would see strong emission lines if such were present
in any galaxy of {\it any} spectral slope with $z < 1.2$
\footnote{At $z > 1.2$, the 3727\AA\ emission line of [OII] is
shifted to 8200\AA, where the night sky emission is becoming
fierce.  At $z > 1.5$, the line is shifted to the point where
the quantum efficiency of
most CCD detectors is rapidly declining.}
if that galaxy fell within our magnitude-limited survey.

\subsection{The UV SEDs of Starburst Galaxies from $z \sim 0$ to $z \sim 3$}

Calzetti, Kinney \& Storchi-Bergmann (1994), in their study of local
starburst galaxies, 
define the UV extinction in terms of
the spectral power law index $\beta$ derived for $f_{\lambda}$, where
in terms of our UV index $\alpha_{UV}$, 
$\beta =\alpha_{UV} - 2.0$.  Their latest calibration of
$\beta$ versus E($B-V$) and versus absorption at 1600\AA\ 
is given in Meurer, Heckman \& Calzetti (1999).  Their data demonstrate
that local starbursts have $-0.5 < \alpha_{UV} < +2.5,$ with a median of
${\sim}0.6$.


There are too few spectroscopically 
identified starbursts in our sample to examine their SED properties.
\footnote{Ignoring the
the starburst galaxy at
$z  = 0.137$, whose UV spectral index is not reliably determined,
the five starbursts spectroscopically 
identified as such in our sample have a very wide range in $\alpha_{UV}$ from
0.1 to 2.7, with a very poorly determined 
median $\alpha_{UV}$ of 0.6.}  Instead we consider the SED properties of
the entire population of $\cal  E$
galaxies.  Figure~8a shows that for the ``mid'' and ``high'' redshift
group (as is true also of the other two redshift groups)  
there is a falloff on the blue
side of the
distribution of $\alpha_{UV}$ at a value of $\sim$0.6. Moreover both Steidel
\etal\ (1999) and Meurer \etal\ (1999) find for $U$-dropouts
at $z \sim 3$ 
about the same median $\beta$ of $-1.6$ (corresponding to 
$\alpha_{UV} = +0.6$)
and deduce from this a median E($B-V) \approx{0.15}$ mag.

While this result was anticipated and is not a surprise, it is by no means
one that is guaranteed.  The mean starburst SED is not
as blue as that of the hottest known stars, but rather represents a sum over
a population including some very hot stars.  There is no reason
beyond our perception of what is logical
why a very young galaxy could not have an even bluer
UV SED, as do a very small number of local galaxies whose UV
light may be dominated by the contribution of a few
Wolf-Rayet stars (Sullivan \etal\ 2000, Brown \etal\ 2000).  
The bluest possible galaxy SED would be the Rayleigh-Jeans tail
of a population of very hot stars, with $\alpha_{UV} = -2.0$, much bluer
than that of observed galaxies.  The mass of the highest mass star
in a young population and the nature of the UV SED
from a starbursts has
been studied theoretically by Elmegreen (2000).  He derives colors
considerably
bluer than observed starburst colors and is forced to introduce 
a mechanism to produce a cutoff in the upper mass below that expected.

We thus
find strong evidence that the most extreme starburst galaxies
(i.e. the bluest galaxies in the rest frame UV) have indistinguishable
SEDs in the rest frame UV over the entire redshift range
$0 < z < 3$.

\subsection{On the Existence of Highly Reddened Starbursts in Our Sample }

We now turn to the nature of the far IR objects observed in
the sub-mm regime by SCUBA
(Blain \etal\ 1999, Smail \etal\ 1999,
Lilly \etal\ 1999, and for the HDF, Hughes \etal\ 1998), and their
relationship to the galaxy populations normally studied 
at optical wavelengths.   We use our SEDs to explore the issue
of whether these galaxies might be some tail of the
optically detected galaxy population with unusually high internal reddening.
Smail \etal\ (1999), Richards (1999)
and Barger, Cowie \& Richards (2000), among others,
have suggested that these far IR sources represent
the tip of a vast iceberg of a separate population
of very dusty galaxies at high $z$ 
in which an enormous amount of star formation is occurring
and which are invisible at optical wavelengths.
Are these SCUBA sources closer to the ULIRGs reviewed 
by Sanders \& Mirabel (1996) 
or to the 
redder of the 
Ly break galaxies already observed at optical wavelengths, as suggested by
Adelberger \& Steidel (2000) ?  Do their presence and numbers
imply that it is ``useless'' to study galaxies in the optical ?

Assuming SCUBA sources are not AGN but rather dusty starbursts,
we attempt to isolate galaxies from our sample
with a high rate of ongoing star formation most comparable to the 
galaxies detected with SCUBA.  We take the group of
galaxies with strong
emission lines, i.e. those classified as $\cal  E$ as the relevant
sample, recognizing that this is a somewhat more diverse group that 
may contain many galaxies ``older'' and with less current
star formation than those detected by SCUBA or than pure
local starburst galaxies. Hence
the intrinsic UV power law index 
of many of these galaxies may be somewhat redder than that of
a pure starburst.

Figure 8a suggests for strong emission line galaxies in our
sample a continuous distribution of $\alpha_{UV}$
peaked at a level slightly redder than that of local starbursts
and of $z \sim 3$ $U$-dropouts.  There is no obvious second (redder)
component, although there may be an extended low level
red tail to the distribution. 
One should, however, note that the distribution in $\alpha_{UV}$ shown
in figure~8a for the $\cal  E$ galaxies includes both $\cal  E$
and $\cal  EI$ galaxies. 
\footnote{Unless otherwise specified, throughout this paper, a galaxy's
spectral class for purposes of plots and figures
is defined only by the first character of its assigned spectral class.}
When restricted to just $\cal  E$ galaxies, 
the red tail
is reduced relative to the blue peak.  Furthermore, galaxies in the
red tail of this distribution do not seem to have lower luminosities,
as would be expected were the red tail due primarily to internal reddening.

This whole issue is related to the nature of the extremely
red objects found in deep infrared selected samples.
As we have stated earlier (Cohen \etal\ 1999a and Cohen \etal\ 2000),
see also Scodeggio \& Silva (2000),
evidence from our redshift surveys suggests that
the majority of the EROs in this magnitude
regime with with $R-K \sim 5$ to 6 
are  passively evolving ``old'' stellar population at $z \sim 1.3$.
While there are
undoubtedly some very red very dusty ULIRGs and with extensive
star formation similar to the ERO HR10 at $z = 1.44$ (Dey \etal\ 1999
and references therein), these appear to be rare.

Our analysis of the SEDs of galaxies with strong emission lines
in our $R$-selected sample
does not provide any evidence that supports
the existence a significant ``missed'' population
of dusty galaxies with extensive star
formation which might be forming
a significant fraction of the stars in 
the Universe within the regime $z \lesssim 1.5$. 
Dust enshrouded
starburst nuclei which do not dominate the total
integrated light from a distant galaxy cannot be excluded by our SEDs.
However, Moriondo, Cimatti \& Daddi (2000) and Stiavelli \& Treu (2000)
find that HST images of EROs  are extremely red not just
within the nucleus but over the entire galaxy image. 

\subsection{Prediction of the Mid-IR Flux from the UV Spectral Index 
\label{ISO} }

Meurer, Heckman, Calzetti (1999) have provided a formalism
for predicting the far IR thermal dust emission from the UV
spectral indices of starbursts.  They suggest that as the
amount of dust increases, the UV spectral index becomes redder
and deviates further from the extremely blue power law index characteristic
of bare young starbursts.  The absorbed UV radiation is then re-emitted
by dust in the mid and far infrared.  Meurer \etal\ (1999) 
give the relevant relationships calibrated
by their analysis of local starbursts to predict the emitted flux 
in the thermal IR from dust.

We attempt to use their formalism to predict which of the 
galaxies in the region of the HDF should have been detected
in the mid-IR by ISO, assuming that a correlation between
mid-IR and far-IR emission exists.  We employ a ranking
scheme to do this, comparing the list of galaxies
actually detected by ISO 
from the analysis of Aussel \etal\ (1999) with our entire 
sample.  The optical 
counterparts of the ISO
galaxies are discussed and listed in Cohen \etal\ (2000). 
Only those with $R < 22.8$ (i.e. those for which the 
identification is considered reasonably certain) are included.
Since the ISO optical counterparts are of 
spectral type $\cal  E$, $\cal  EI$ or $\cal  EA$ (Cohen \etal\ 2000), 
we also predict
the mid-IR flux for the galaxies within our sample of those
spectral types (as well the
the spectroscopically identified starbursts, which are mostly
of too low luminosity to
have been detected by ISO) and with $R < 22.8$. 
 
Because of the correlation between galaxy spectral type and SED indices
shown in Tables~2,3 and in Figure~8, a mis-classification of a galaxy
to spectral type $\cal  E$ which is actually an intermediate spectral
type ($\cal  I$, $\cal  EI$ or $\cal  IE$)
will produce a galaxy that appears slightly too red in the rest frame UV, 
hence seems excessively dusty. The predicted mid-IR flux
for such a mis-classified object will be overestimated.
This type of classification error should not occur until a galaxy
is either very faint or has 
$z > 0.9$, where the 4000\AA\ region
is shifted into the thicket of night sky lines and it is
very hard to distinguish anything except the emission features.
Hence the mid-IR luminosity prediction is 
carried out for the redshift range 0.25 to 1.05, omitting
the highest $z$ range, where the galaxy spectral types
are more uncertain.

There are 12 ISO galaxies with secure optical counterparts
in the relevant redshift range.
We find the ranking of the 12 ISO galaxies in predicted mid-IR
luminosity using the formalism of Muerer \etal\ 
within the set of 154 potential optical counterparts
from our survey selected as described above.  
For comparison, we ascertain the rank of the ISO galaxies
in the total sample when ordered by $L(B)$ or by $L(R)$. 

Figure~10 shows the rankings of the 12 ISO detected galaxies
in this redshift range within the sample of 154 galaxies.
A perfect discriminant would have the ISO galaxies occupying ranks
1 through 12.  None of the discriminants tried is very good.
The results are unchanged when only $\cal  E$ galaxies are used.
They are also unchanged when a correction to the luminosity for passive 
evolution is made.  The
sBB SED model fits were used throughout; the results are unchanged
when the 2p fits are used.

The best discriminant appears to be the $B$-band luminosity.
The mid-IR flux predicted with the algorithm
of Meurer \etal\ (1999) for
the galaxies detected by ISO in our sample does not
appear to do better overall than either the
$B$ or $R$ band luminosity.
While the ISO detections
correspond to galaxies which are among the most luminous (using
any of the three definitions of luminosity given above), there
are still many galaxies of comparable luminosity that were not
detected by ISO.  The formalism of Meurer \etal\ (1999)
does not significantly reduce this problem.  One should note, 
however, that this formalism was developed to explain 
the far-IR emission of galaxies where 
thermal emission from dust dominates.
At a typical redshift of 0.6, the 15$\mu$ ISO bandpass
corresponds to 9.4$\mu$ in the rest frame.
The majority of the ISO detections were in the 7$\mu$ band,
where, in the rest frame for these distant galaxies, 
dust emission does not dominate the total emitted galaxy light.

\section{Comments on Photometric Redshifts \label{photoz}}

Photometric redshifts have become very popular recently; see, for
example, Connolly \etal\ (1997) or Fernandez-Soto, Lanzetta \& Yahil (1999).
As our blind tests (Hogg \etal\ 1998, Cohen \etal\ 2000)
have shown, these are capable of accurately predicting the true redshift
of an object in essentially all cases  
with the excellent photometry that one can obtain from deep HST
images.  But many wide area ground based surveys plan
to use photometric redshifts for various purposes, and one might be
concerned that the limitations of ground based photometry
for such faint objects will reduce the accuracy of the resulting
photometric redshifts beyond a tolerable level.  

There are two problems here, both becoming worse as one presses fainter.
The first is the impact of crowding 
while the second is the decreasing accuracy of the photometry itself for
isolated fainter objects.  
We can assess the influence of the former in that even with very 
careful hand checking of all objects in our sample in
the region of the HDF and with images with 1 arcsec
seeing, we could not obtain believable
SEDs for about 25 of 590 objects because of crowding.
\footnote{The number of galaxies without reliable SEDs
given in \S\ref{problems} includes galaxies with no 
observation at $K$ as well as those suffering from
crowding.}
Almost all
of these objects have $R > 22$. That error
rate in a massive non-manually checked catalog at this magnitude
level would be larger and  would further increase
by a large factor for a catalog with a fainter limiting
magnitude or  poorer seeing conditions.
The influence of confusion on photometry is discussed by Hogg (2000).
Even in the HDF itself, our survey still operates at a level in excess of
60 resolution elements per source.

The 4000\AA\ break is the dominant feature in the spectra of galaxies
in this redshift range. 
\footnote{We ignore the higher redshift range where the Ly break dominates
and where the decline in
flux across the boundary is more than a factor of 10, but the
objects are in general fainter than the range considered here
of $R < 24$.}  If the photometric errors are such that
it is impossible to discern the presence of the break, assignment
of a photo-$z$ becomes problematic at best.

We can test with each of our two fitting models whether
a 4000 \AA\ break is measurable.  
In the case of the 2p SED model, we require that
we can detect  
$\alpha_{IR} - \alpha_{UV}$, which is a broad band change
in continnuum slope, as non-zero, while the sBB model
permits direct measurement of the 4000 \AA\ break.

Our criterion  for detectability for the 4000 \AA\ break in
the sBB SED model is that  
$L(\lambda_m,\rm{red}) - L(\lambda_m,\rm{blue})
\ge 0.05$ dex (corresponding to a 4000 \AA\ break of $\ge$0.13 mag).
For the 2p model, we require that 
$|(\alpha_{IR} - \alpha_{UV})|/\sigma \ge 1.5$, where
$\sigma$ is determined from the fitting procedure and is
based on the observed brightness
of the object, the number of filters used to
determine the SED, and the redshift of the object.
 
We find the number of galaxies in each redshift range
for which, with the definitions given above,
the detection of a 4000 \AA\ break is problematical.
This fraction is less than 20\% for galaxies with
$z < 0.8$, but rises to more than 30\% above $z\sim0.8$. 
While the overall fraction of galaxies without detectable
4000 \AA\ breaks is approximately the same for the two SED models,
the behavior of galaxies of spectral class $\cal  A$,
in particular, depends on the fitting scheme adopted.  
In the 2p model, essentially all such galaxies have
easily detectable breaks, while in the more detailed sBB model, they have
small 4000 \AA\ break, but still display substantial changes in overall
spectral slope between the rest frame UV and optical.
In any case, there is clearly a significant 
population of galaxies, consisting predominantly of
those with strong emission lines, that do not have
a detectable 4000\AA\ break or Balmer jump.  For such galaxies,
adding IR photometry will not help in providing a valid photometric
redshift.  Only higher precision photometry, difficult to achieve 
from the ground for such faint objects, will help.

Brunner, Szalay \& Connolly (2000) have discussed this issue 
at length as 
it affects their study of galaxy clustering, which uses photometric
redshifts to divide a pencil beam survey into several redshift
shells, thus avoiding the projection integral. Their particular
application is eased by the fact that, as shown by
Hogg, Cohen \& Blandford (2000), throughout this range in $z$
most of the clustering signal is from
the redder early type galaxies, which will yield reasonably
accurate photo-$z$s under any appropriate scheme.

\section{A Qualitative View of the Evolution of $L^*(K)$ \label{klum}}

We show in Figure~11 the rest frame $K$-band luminosity $L(K)$ 
calculated from the sBB parameters of the SED fits
as a function of cosmological comoving volume.
The issue of K-corrections has
been avoided through the use of multi-color photometry spanning
a broad wavelength range.  However, extrapolation beyond  the
reddest observations (i.e. trusting the functional form of the
model SEDs) is required to
obtain rest frame $K$ magnitudes for galaxies in this sample, with
the largest extrapolation required for the highest $z$ galaxies.

The rationale for studying galaxies at $K$ is that
the integrated light there is much more representative 
of the total stellar mass of a galaxy than are the 
optical colors, where light from the most recent epoch
of star formation may dominate over that from the
older population as one moves towards the ultraviolet.
To convert these $K$ luminosities for the galaxies in
our sample into total
stellar masses of the galaxies requires a model of evolving
galaxy spectral energy distributions
to evaluate their mass-to-light ratio as a function of look back time
and of their star formation history.
We use the evolutionary corrections from the
models of Poggianti (1997), and interpret her
E, Sa and Sc galaxy classes as roughly equivalent
to our galaxy spectral types $\cal  A$, $\cal  I$ and $\cal  E$.
The lines in Figure~11 thus indicate the predicted
track of a galaxy of constant mass from Poggianti for the
three mophological galaxy classes.
As expected, the evolutionary
change in luminosity at $K$ predicted by Poggianti is quite
small and much less dependent on the details
of the star formation history (i.e. of the galaxy spectral
class) than are those predicted at $B$ or $R$.  These tracks, which
include passive stellar evolution for the elliptical, are at rest frame
$K$ a good representation of the evolution of at least the most luminous 
galaxies in our sample.

Recent determinations of $L^*(K)$ in the local Universe correspond
to $L^*(K) = 8 {\rm\times} 10^{36}$ W (Gardner \etal\ 1997,
Loveday 1999, and from the combination of the 2dF and 2MASS survey,
Cole \etal\ 2000).  The passive evolution
model lines are shown with this luminosity (at $z=0$), equivalent
to $1.3 {\rm\times} 10^{11} L\subsun$ at $K$.  For elliptical
galaxies, using the 
mass-to-light ratio at $K$ for local ellipticals computed by Worthey (1994),
this is a galaxy with a total mass of $1 \times 10^{11} M\subsun$.
The mass-to-light ratio of Sb galaxies is smaller, by about
a factor of 2.

Figure~11 suggests, as has been evident from our work on
spatial clustering in this field, that luminous nearby
($z \lesssim 0.3$) galaxies are absent in this field.  That is
presumably a result of the selection criteria for
defining the HDF adopted by Williams \etal\ (1996).

Figure~11 also shows the gradual increasing mean luminosity of
star forming galaxies (our spectral class $\cal  E$) to 
$z \sim 1$, and the increasing dominance of such
galaxies as a fraction of the total observed population, an effect found
previously with smaller samples by many groups, including 
Cowie \etal\ (1996), Hammer \etal\ (1997) and Hogg \etal\ (1998).
There are some selection effects, as discussed many
times in the earlier papers in this series, namely faint objects
have a lower probability of being assigned a redshift unless they
have very strong emission lines and 
the same is true for galaxies with $z > 1$.  However, the
redshift completeness of our sample is very high, in excess of 93\%.

A formal analysis of the galaxy luminosity functions will follow
in the next paper in this series.  Once the evolution of
$L^*$ is properly determined, then
subject to the accuracy of Poggianti's (1997) predicted
evolutionary corrections at rest frame $K$ and to the
uncertainties in the extrapolation of our SEDs beyond the range
in wavelength over which photometry exists, one can evaluate
the change in the total stellar
mass of $L^*$ galaxies. This can then be used to provide another constraint
on the rate of massive mergers
out to $z \sim 1$; Carlberg \etal\ (2000) has already provided
one such constraint through an analysis of
the kinematic pairs in our sample.

\section{Summary}

In this paper we have evaluated the spectral energy distributions for
a sample of 590 galaxies in the region of the HDF with $z < 1.5$ using
a new versatile and stable SED model.  This
has been done directly from the data using multi-color photometry
with wide wavelength coverage combined with spectroscopic redshifts
from our 93\% complete $R$-selected redshift survey here.
The behavior of the
SEDs with galaxy spectral type (i.e. ``recent'' star formation history) 
and with redshift confirms the trends found in
our earlier studies of smaller samples; galaxies with strong signs of
recent star formation (i.e. those which show emission lines) have 
bluer continuum slopes in both the rest frame UV and the optical/near-infrared.
The most luminous galaxies tend to be of galaxy spectral class $\cal A$
at a rate much higher than their fraction in the total sample, i.e.
the redder galaxies tend to be more luminous.
In the mean, actively
star forming galaxies become bluer in the rest-frame UV at higher
redshifts, which may perhaps be due to systematic redshift dependent
fitting errors for $\alpha(UV)$.  We see no other
change with redshift in the relationship between SED characteristics
and galaxy spectral type defined by the presence and strength
of narrow emission and absorption features  
to $z \sim 1.1$.   

We use these SEDs to evaluate the potential accuracy of photometric
redshifts, bearing in mind that a break at 4000\AA\ must be detectable
to within the errors of the photometry to assign a photo-$z$ for
galaxies in this redshift regime.  We also use them to 
demonstrate that the SEDs in the rest frame UV of the 
the most extreme starburst galaxies (i.e. the bluest 
$\cal  E$ galaxies) are indistinguishable SEDs 
from those of local starbursts as analyzed by Calzetti, Kinney
\& Storchi-Bergmann (1994) and from those of
the Ly break galaxies at $z \sim3$ studied
by Steidel \etal\ (1999) and by Meurer \etal\ (1999).

We also  attempt to set constraints on the possible
existence of a separate class of dusty starburst galaxies and on the
variation from galaxy to galaxy within a galaxy spectral class of
internal reddening using these SED indices.
Finally we use
the UV spectral indices of all the strong emission line galaxies
to predict which of them should have been detected by ISO in the
mid-IR, which exercise was only modestly successful.

We conclude by presenting
the rest frame $K$-band luminosity as a function of $z$.  
We see the effect of selecting a field (i.e. the HDF)
to be devoid of bright galaxies.
We see encouraging 
overall consistency with predictions of evolutionary corrections
for the rest frame $L(K)$ computed from models of integrated 
light of galaxies by
Poggianti (1997).  We see the gradual change in the population
of various types of galaxies, with star forming galaxies becoming
both a larger fraction of the total population 
and more luminous as one moves toward $z \sim 1$.
The overall pattern of the $L(K) - z$ relationship
suggests that passive
evolution at constant total stellar mass is a good approximation to the actual
behavior  of at least the most luminous galaxies in
this large sample of galaxies in the region of the HDF
out to $z \sim 1.5$, an issue which will be explored in depth
in the next paper in this series.

\acknowledgements The entire Keck/LRIS user community owes a huge debt
to Jerry Nelson, Gerry Smith, Bev Oke, and many other people who have
worked to make the Keck Telescope and LRIS a reality.  We are grateful
to the W. M. Keck Foundation, and particularly its late president,
Howard Keck, for the vision to fund the construction of the W. M. Keck
Observatory.  

We thank Roger Blandford, David Hogg and Gerry Neugebauer 
for helpful discussions.
We thank Amy Barger and Len Cowie for access to their unpublished photometric
database for the region of the HDF.  We thank Robert Brunner for
supplying filter transmission curves.
We thank two anonymous referees for constructive criticism of the manuscript.

This work was not supported by any federal agency.

\appendix
\section{Updates to the Redshift Catalog in the Region of the HDF}

We present in Tables 6a,b redshifts obtained since
November 1999 for objects in the region of the HDF.   This
is new, additional material which supplements that presented 
in Cohen \etal\ (2000). The galaxy spectral classes and redshift
quality classes used here
are those defined in Cohen \etal\ (1999a).  Table~6a
contains information for 53 objects in the Flanking Fields and is
based exclusively on our LRIS (Oke \etal\ 1995) observations at the 
Keck Observatory during
the winter of 2000.  Emphasis in planning the observations
was on galaxies with $23.0 < R < 23.5$.

Table~6b contains the additional redshifts for five objects in the HDF
itself, some of which is from other groups as indicated in the table.

Combining this new material with the data presented in Cohen \etal\ (2000),
the redshift completeness in the HDF itself to $R < 24$ is now 95\%,
while that in the Flanking Fields around to within the survey area
of Cohen \etal\ is in excess of 93\% to $R < 23$.

The SED parameters for these new galaxies with $z < 1.5$ 
for which reliable SEDs are given in Table~7a,b.

In addition, three redshifts from the catalog of Cohen \etal\ (2000)
are being changed. The spectral features
which led to the redshift assignment given in our catalog in each case are 
real, but the
identification assigned to them has been modified.
Based on a suggestion from Mark Dickinson (private communication),
the same suggestion subsequently being made by 
Fernandez-Soto \etal\ (2000), and after
looking at the relevant spectra,  the redshift of
H36396\_1230 has been changed. What in hindsight should have been recognized
as an obvious Ly$\alpha$ emission line was previously identified 
as the Mg II line at 2800\AA.  The redshifts of 
H36494\_1316 and H36560\_1329 are also being modified to identify
the single emission line as [OII] 3727\AA\ instead of H$\alpha$ 
to be in conformance with the
rules stated in earlier papers in this series regarding
redshifts from spectra showing only single emission lines 
and also based on the
photometric redshift/spectroscopic redshift discrepancy discussed
by Fernandez-Soto \etal.  If these rules had been followed
in all 3 cases, no modifications would have been necessary.  The
corrections are listed in Table~8.  The SEDs have been calculated using
the updated values.

Finally, the magnitude given for H36453\_1153 in Cohen \etal\ (2000)
is too bright.  There are several faint objects close together
which are clearly resolved in the HST image, but
only one entry appears in the H00 photometric catalog.
The object with the
spectroscopic redshift is the $U$-dropout, and instead of $R = 22.53$,
the value given in Cohen \etal\ (2000), 
a better estimate for this galaxy is $R \sim 23.3$.

\section{Effective Wavelengths of the Filters}

The effective wavelength  of the 
$K_s$ (short $K$) filter is taken as 2.17$\mu$ with an absolute
flux at 0 mag of log[$\nu f_{\nu}$] = $-$9.01 W m$^{-2}$ 
(M.Pahre, private communication). 
Persson \etal\ (1998) also discuss the effective wavelengths of various
infrared photometric systems.  The effective
wavelengths of  the other three filters in the H00 database
($U_n$, $G_n$ and $R$) are given by Steidel \& Hamilton (1993).
The Barger \etal\ (1998) database consists of at least
six colors, five of which are
on the Johnson system, and hence their effective wavelengths
can be taken from standard sources; we adopt
those of Fukugita,  Shimasaku \& Ichikawa (1995).
The notched H$K'$ filter is described in Barger \etal\ (1998); see also
Wainscoat \& Cowie (1992). The galaxy measurements at
$HK'$ were transformed to a standard
Johnson $K$ using the relationship $K = HK' - 0.3$ given
by Barger \etal\ (1998).

\section{Details of the SED Fitting Procedures}

Here we describe the details of the fitting procedures, and how
the flux at the transition wavelength is handled.

For the 2p SED, the power law was fit first to the
side (blue or red) of rest frame
4000\AA\ which had the most observed points;
this depended on the redshift of the object and the number of
filters for which data actually exists.
A linear least squares fitting routine from Press \etal\ (1986) was used.
Once one index was determined,  the second spectral index is calculated
assuming forced agreement in the value of $L_{\nu}$ at 4000\AA.

For the sBB SED model,
the fitting is accomplished with routines from Press \etal (1986).
The fitting rules are as follows.  If there are 
4 or more filter bands observed on the red side and 3 or more on the 
blue side, then the regimes $\lambda < \lambda_m$ and $\lambda > \lambda_m$
are fit independently.  If there are fewer than four on
the red side, then the blue side is fit first. A point is then
added to the red side at $\lambda_m$ with the
luminosity there that calculated from the blue fit.  Then the red fit is done.
If there are fewer than three on the blue side, then the
red side fit is done first, and a point is added to the blue
side at $\lambda_m$ with the value predicted from the
sBB fit.

\clearpage

\begin{figure}
\epsscale{0.7}
\plottwo{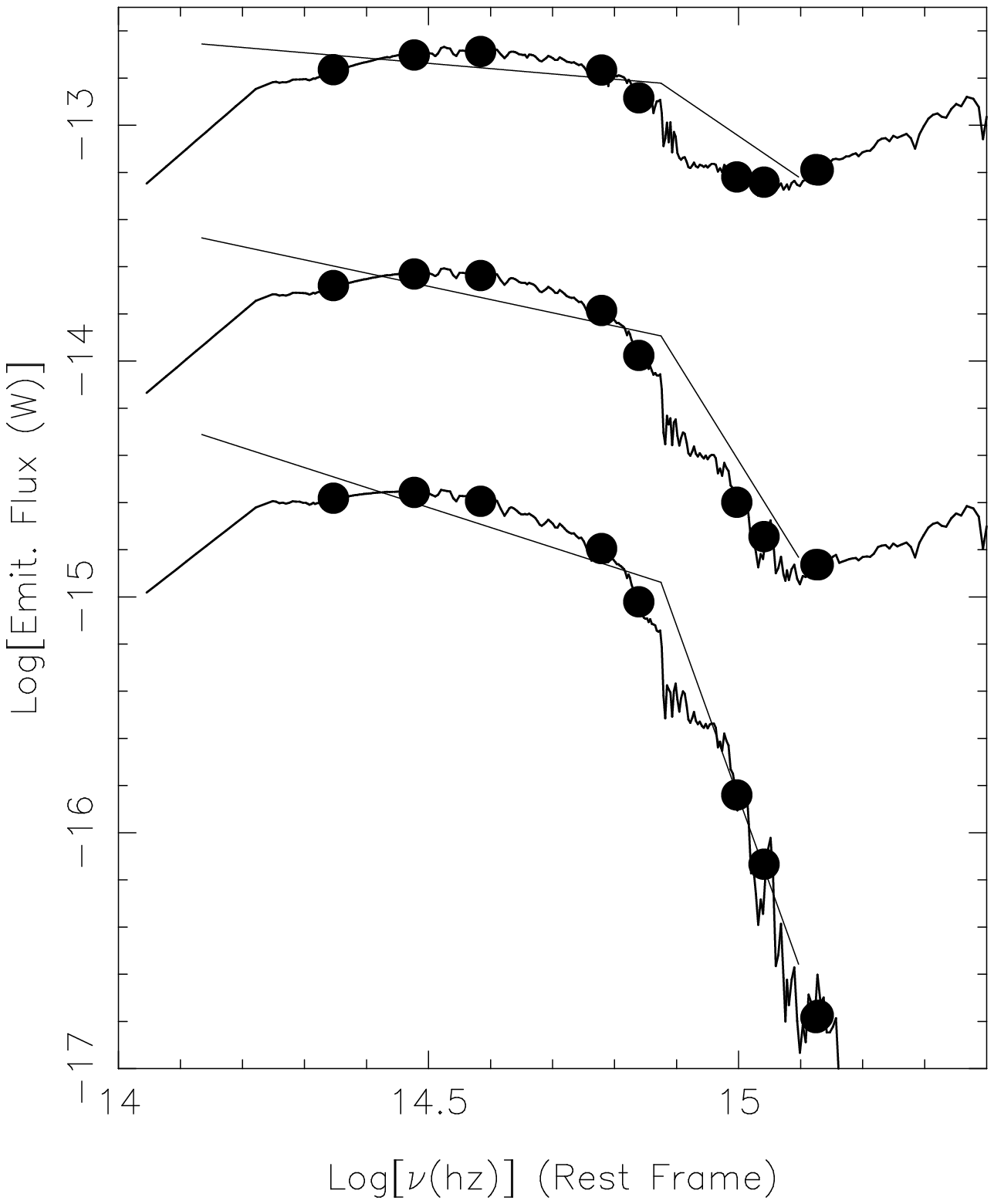}{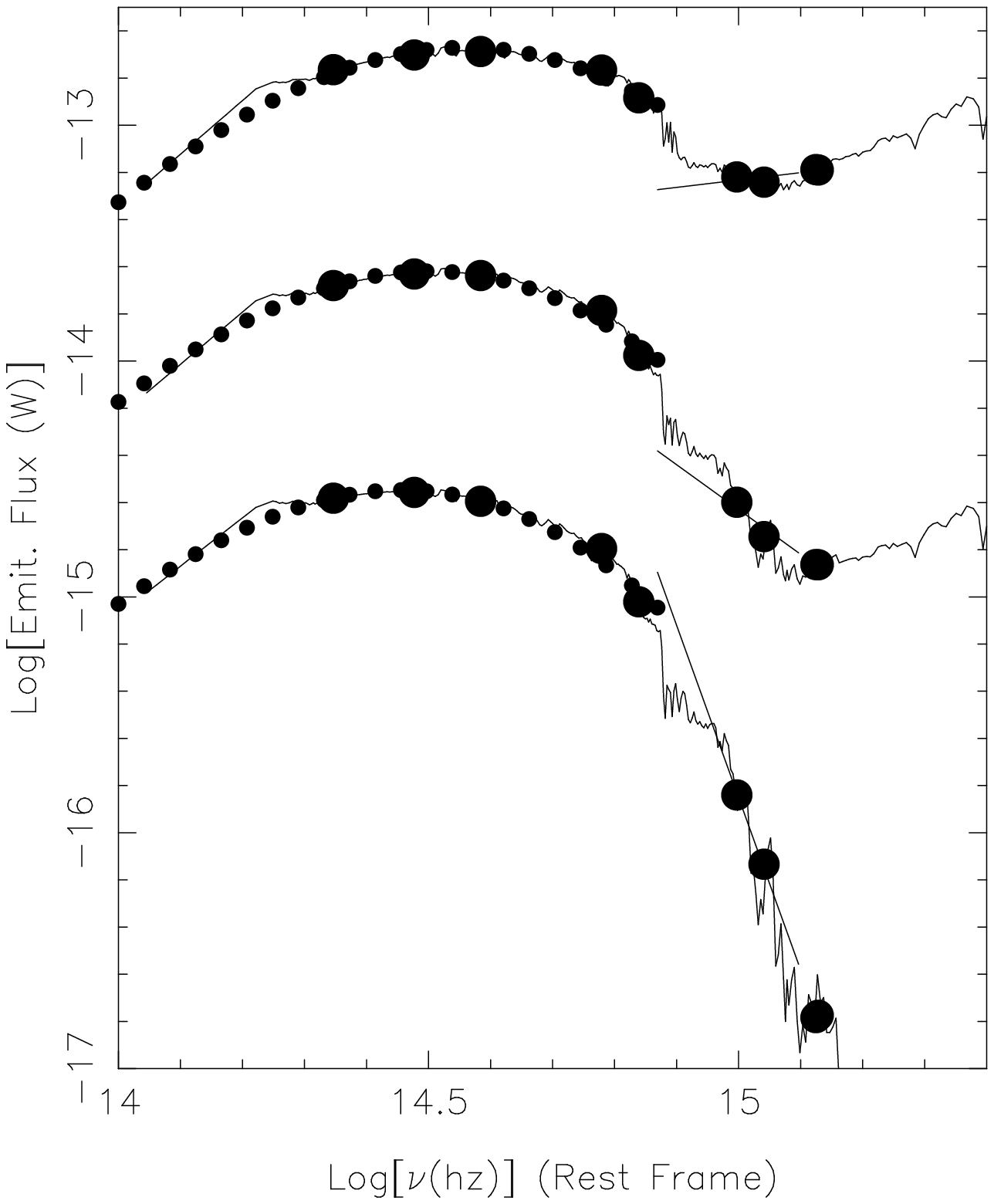}
\caption[figure1a.ps]{a) The SEDs of $z=0$ elliptical, Sa and Sc galaxies from
Poggianti (1997), shown as the curves, are observed at a redshift of 0.6. 
The resulting filter bands
are indicated by the large filled circles. The 2p SED model
is then fit to this set of filter data and the UV and IR fit for each of
the three SEDs is shown as a straight line. b) The same as Figure 1a, but
using the sBB SED model.  The best fit is shown as the small filled
circles for the sBB function and as the solid line for the power law
in the UV.
\label{fig1}}
\end{figure}

\begin{figure}
\epsscale{0.9}
\plottwo{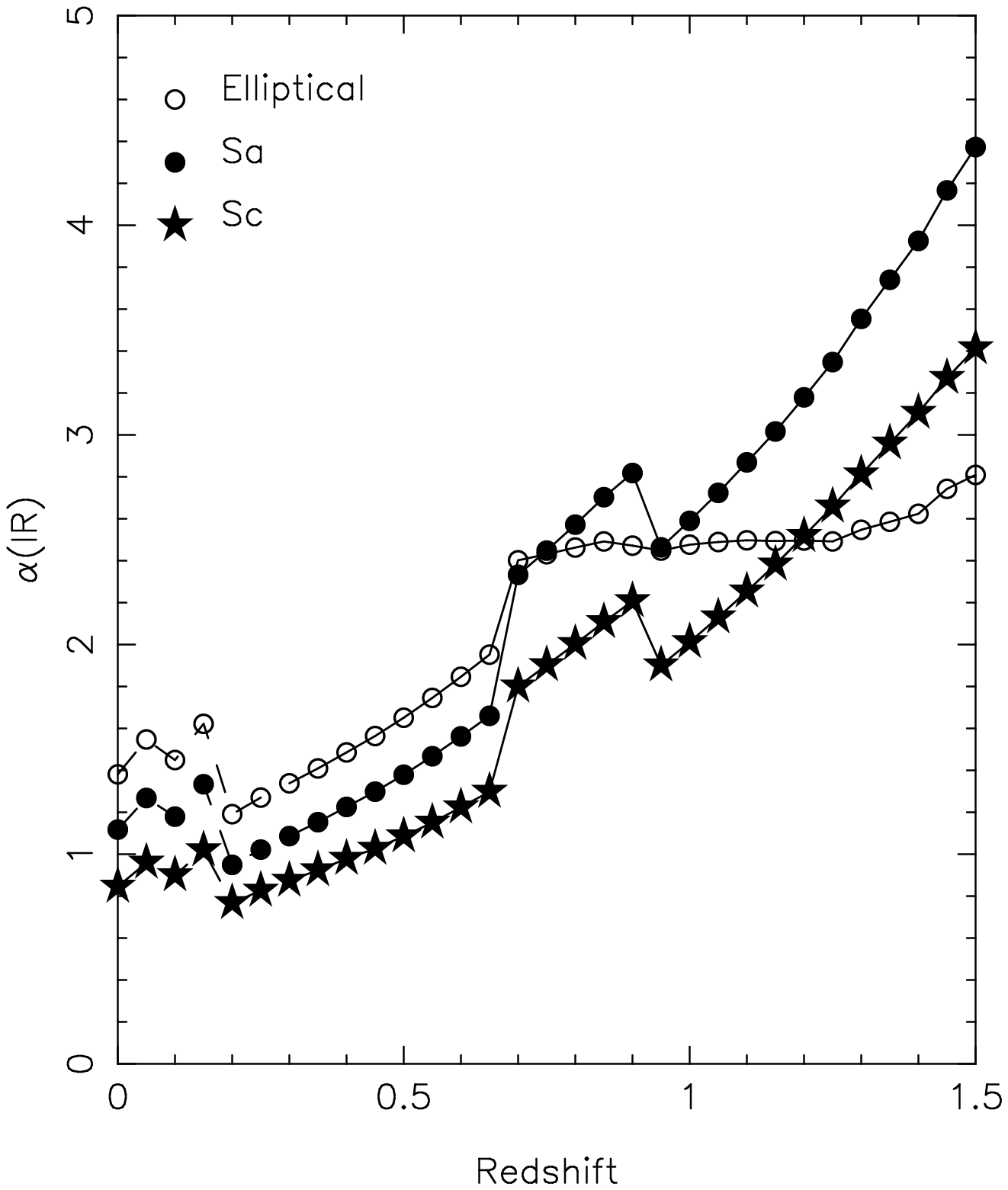}{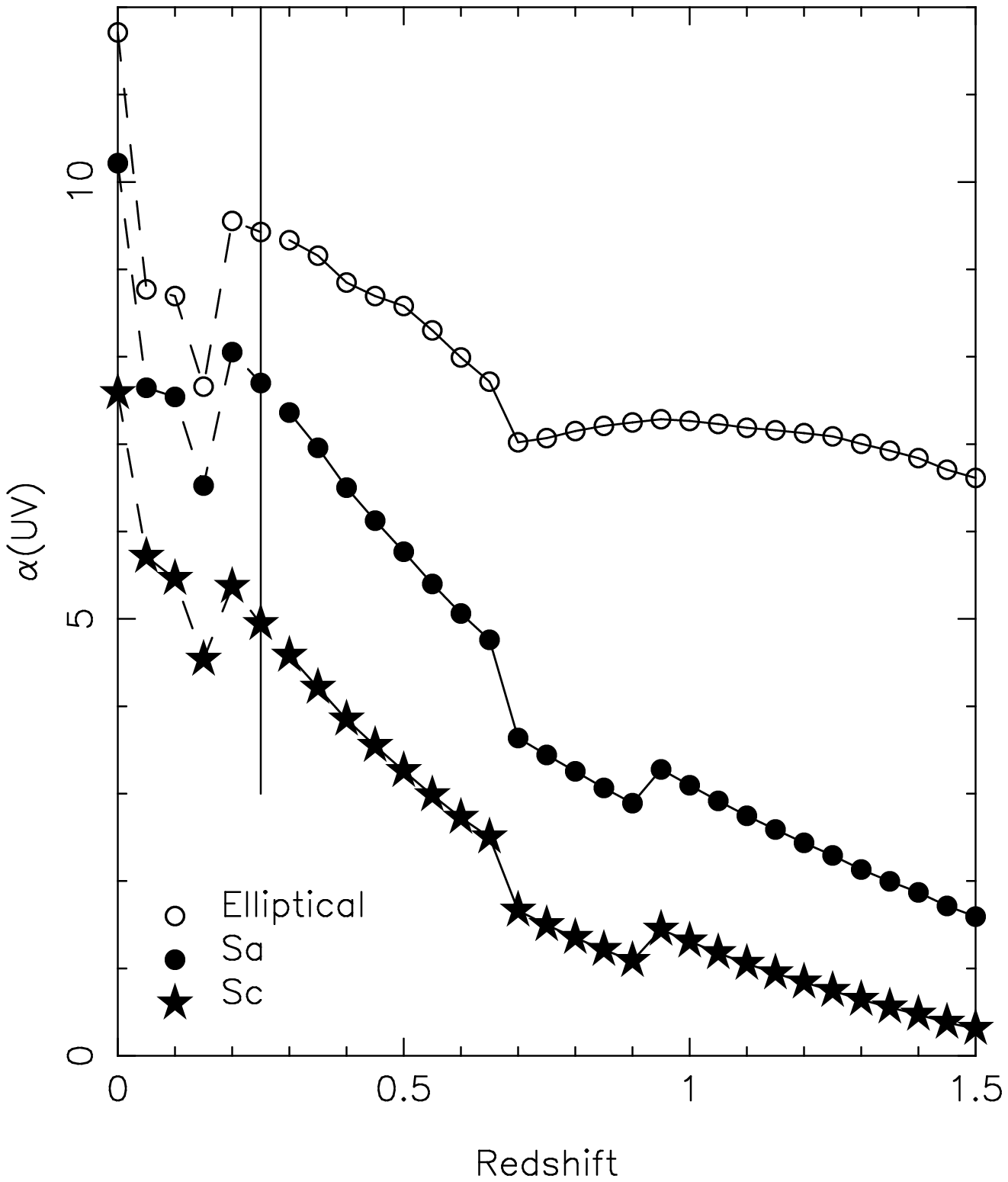}
\caption[figure2a.ps]{The UV and IR power law indices found from the
2p model of galaxy SEDs are shown for the standard test set of 
local SEDs (those of Figure~1).
The  redshift of observation ranged from $z=0.0$ to 1.5.  
Open circles denote the track
of the SED parameters determined for
the local elliptical galaxy SED, filled circles show that of a local Sa, and
stars the local Sc galaxy.
The vertical line in panel (a) reminds us that the UV fit is not
considered valid in the region $z < 0.25$. 
\label{fig2}}
\end{figure}

\begin{figure}
\epsscale{0.9}
\plotone{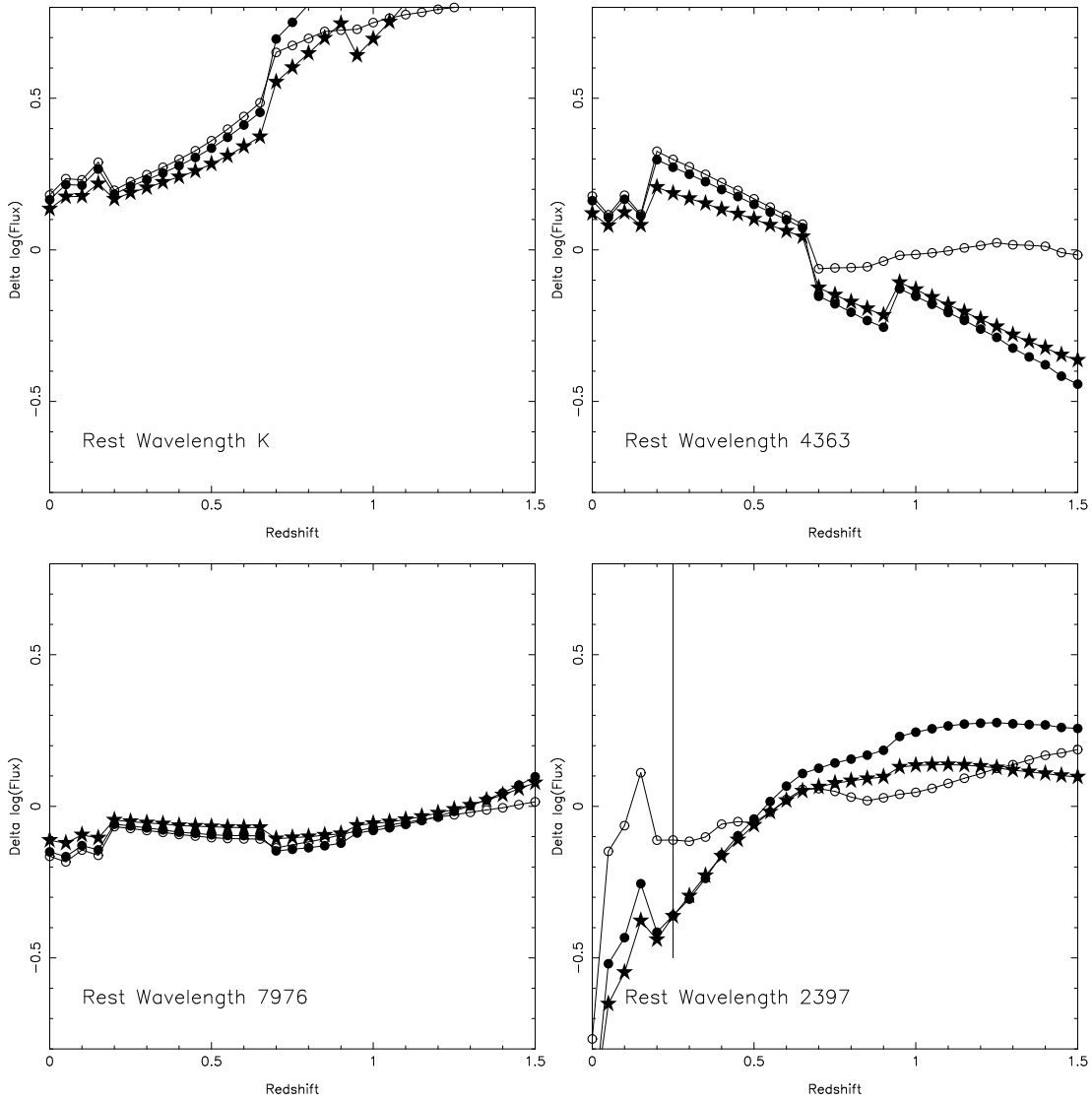}
\caption[figure3.ps]{The rest frame galaxy luminosity
$L(\nu_0)$ predicted by the best fit
2p SED model is compared with the actual luminosity of
the standard test set of local galaxy SEDs, which 
are observed over the range in redshift from 0 to 1.5.
The plot symbols are the same as
those of figure 2.  The four panels give the results for four
different rest wavelengths.
\label{fig3}}
\end{figure}

\begin{figure}
\epsscale{0.7}
\plottwo{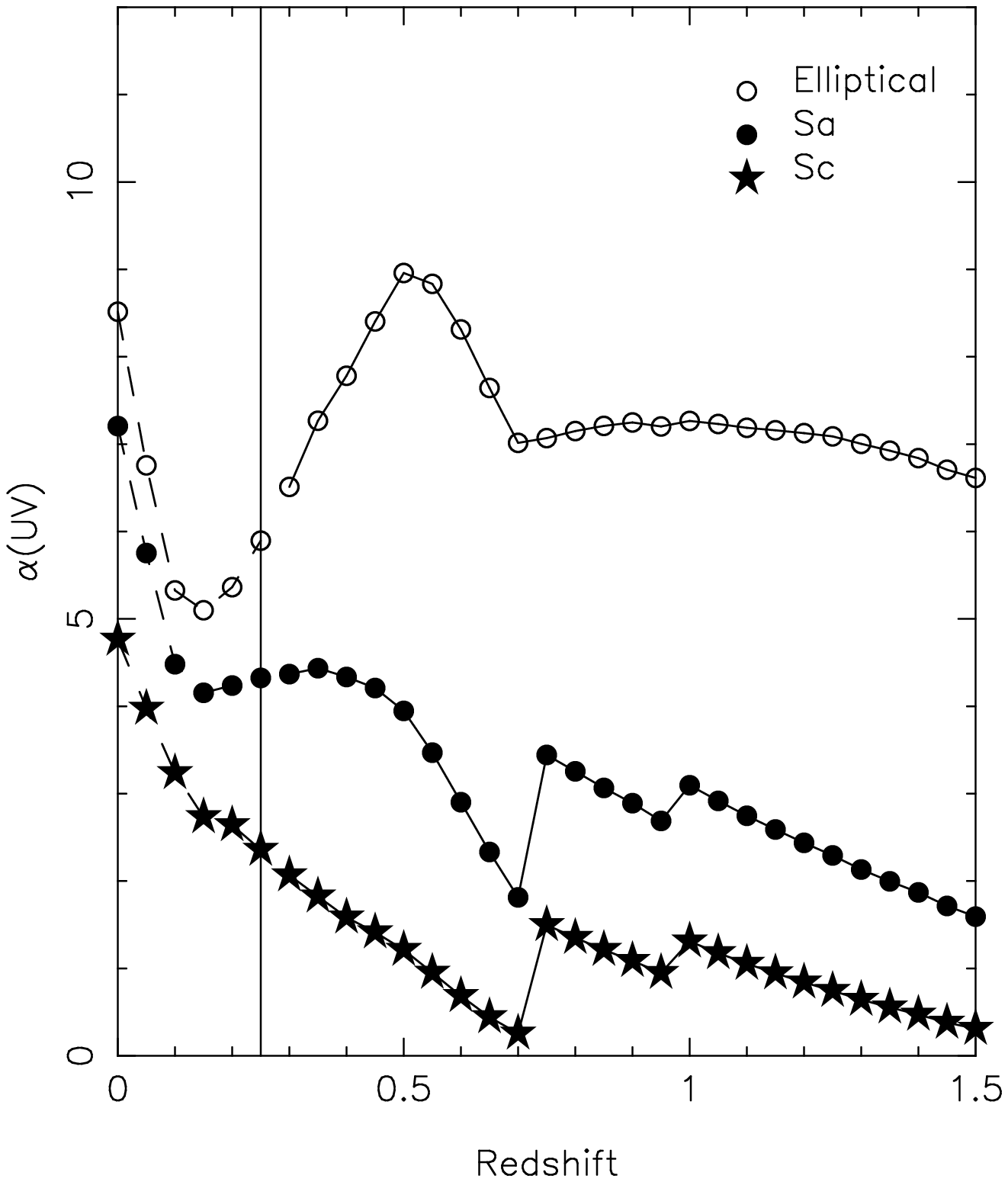}{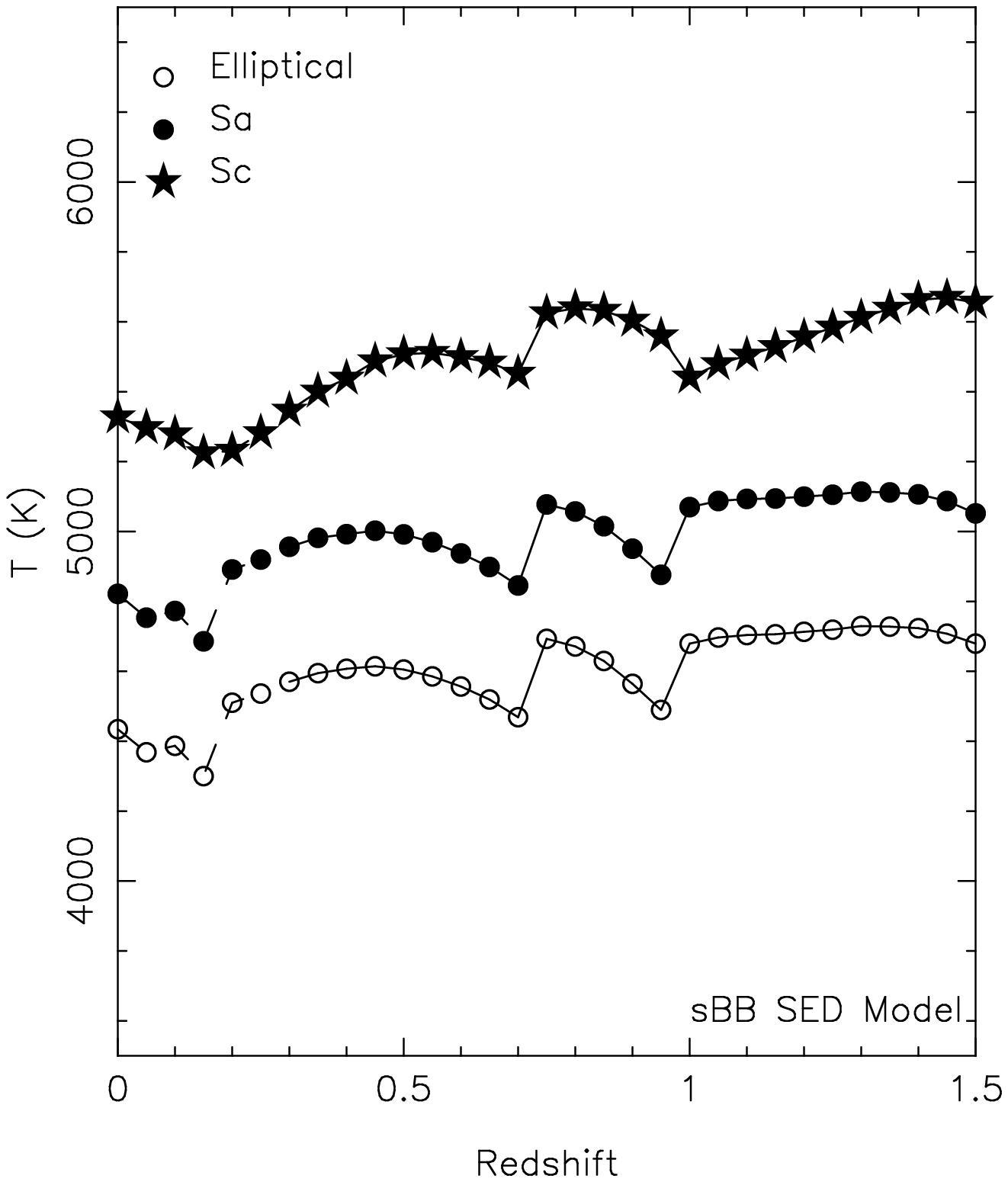}
\caption[figure4a.ps]{The measured parameters from the sBB model 
fit are shown ($\alpha_{UV}$ in (a) and 
$T$ for the optical and near IR in (b))
for the standard test set of local galaxy 
SEDs (those of Figure~1) observed over the
range in redshift from 0.0 to 1.5.  The  plot symbols are the same
as those of figure 2.  The vertical line in panel (a) 
reminds us that the UV fit is not
considered valid in the region $z < 0.25$. 
\label{fig4}}
\end{figure}

\begin{figure}
\epsscale{0.9}
\plotone{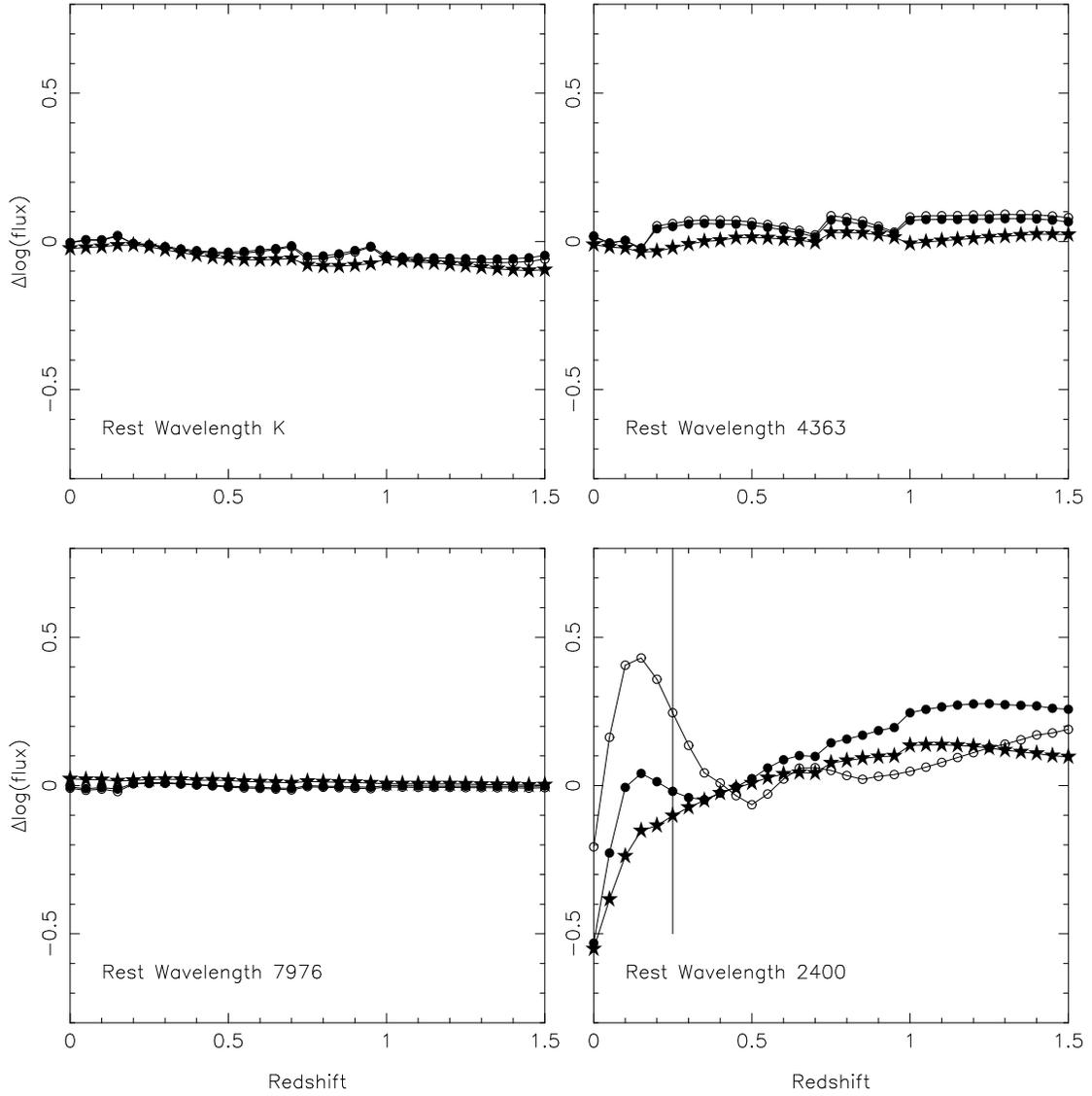}
\caption[figure5.ps]{The rest frame galaxy luminosity
$L(\nu_0)$ predicted by the best fit
sBB SED model is compared with the actual luminosity of the galaxy SED for
the set of standard local galaxy SED models
observed over the range in redshift from 0 to 1.5.
The plot symbols are the same as
those of figure 3.  The four panels give the results for four
different rest wavelengths and are identical in wavelength and scale to the
panels of Figure~3.
\label{fig5}}
\end{figure}

\begin{figure}
\epsscale{0.7}
\plotone{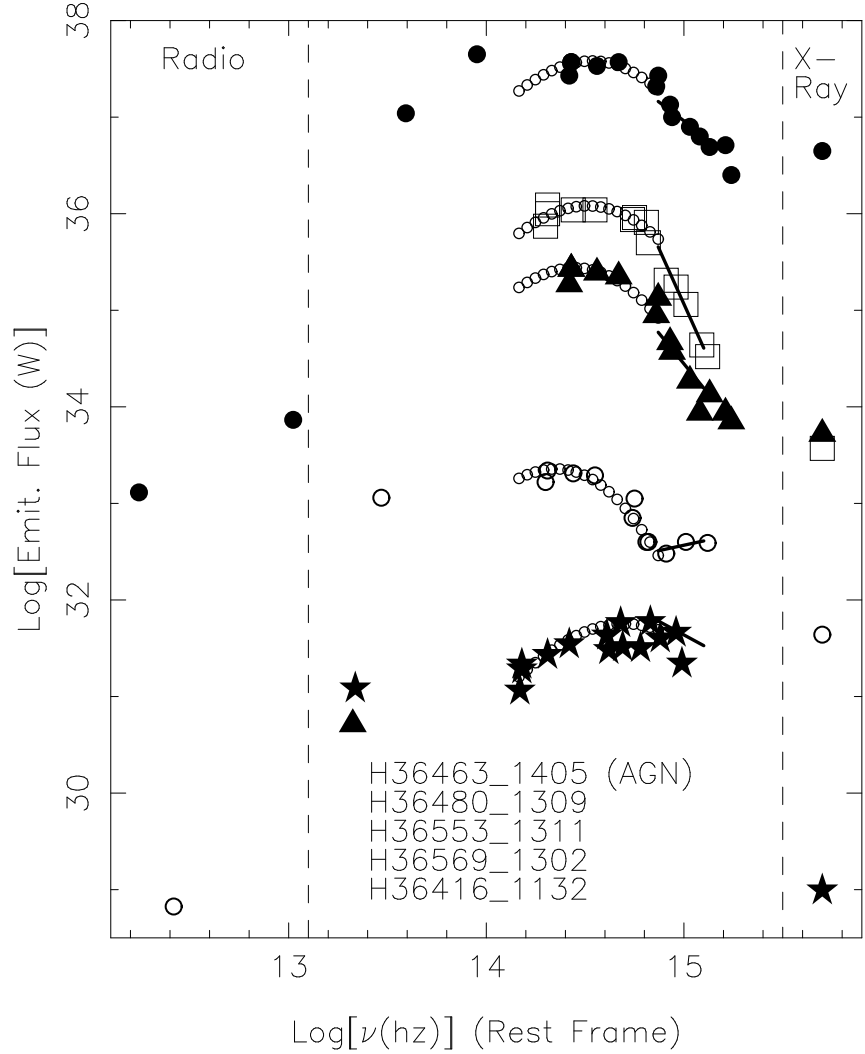}
\caption[figure6.ps]{The SEDs for the five galaxies in the HDF
in our sample
with detections in the X-ray by Chandra from 
Hornschemeir \etal\ (2000) included.  The sBB fits are shown
as well as the mid-IR
detections from ISO from Aussel \etal\ (1999)
and the VLA radio detections  of Richards \etal\ (1998) and Richards (2000)
at 1.4 and 8.5 Ghz.
The frequency scale is correct for
the optical and mid-IR, but is discontinuous at each end to include
the radio and X-ray detections.  The vertical scale is correct
for the uppermost galaxy (H36463\_1405) shown.  A vertical shift of 1 dex 
downward is applied for each additional galaxy plotted.  The vertical
order of the galaxies in the optical is that given in the text
insert at the bottom of the figure.   
\label{fig6}}
\end{figure}

\begin{figure}
\epsscale{1.0}
\plotone{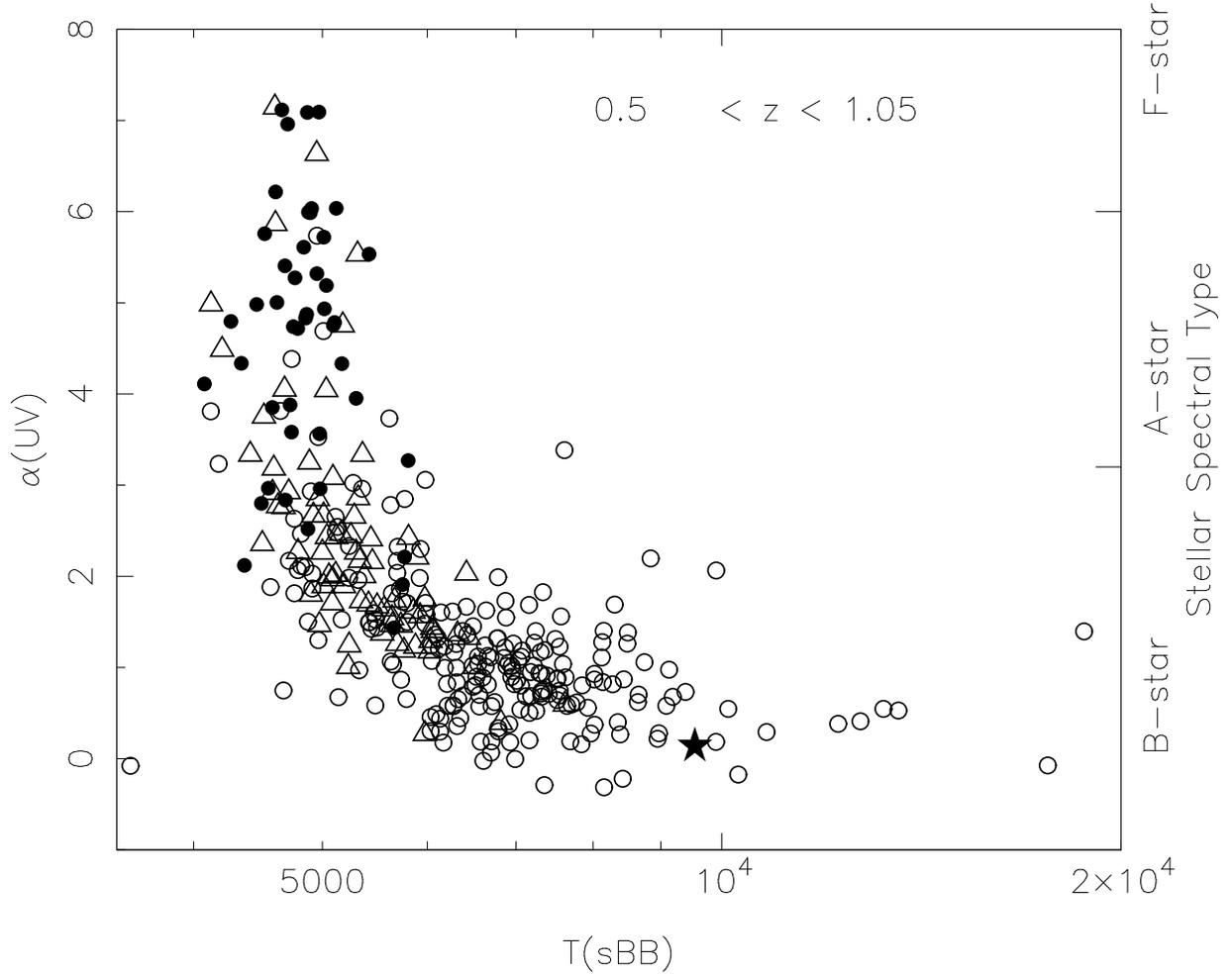}
\caption[figure7.ps]{The spectral indices
$\alpha_{UV}$ are shown as a function of $log(T)$ 
for galaxies in the region of the HDF in the ``mid'' and
``high'' redshift ranges.
Filled circles denote
galaxies of spectral type $\cal  A$, triangles denote
galaxies of spectral type $\cal  I$, and open circles denote
galaxies with strong emission lines (spectral type $\cal  E$).
Starburbursts (spectral type $\cal  B$) are denoted by stars.
Note that only the first letter of the galaxy spectral class is used.
The range of $\alpha_{UV}$ for stars of various spectral types,
taken from Cohen \etal\ (1998a),  is
shown on the right side of the figure.
\label{fig7}}
\end{figure}

\begin{figure}
\epsscale{0.9}
\plottwo{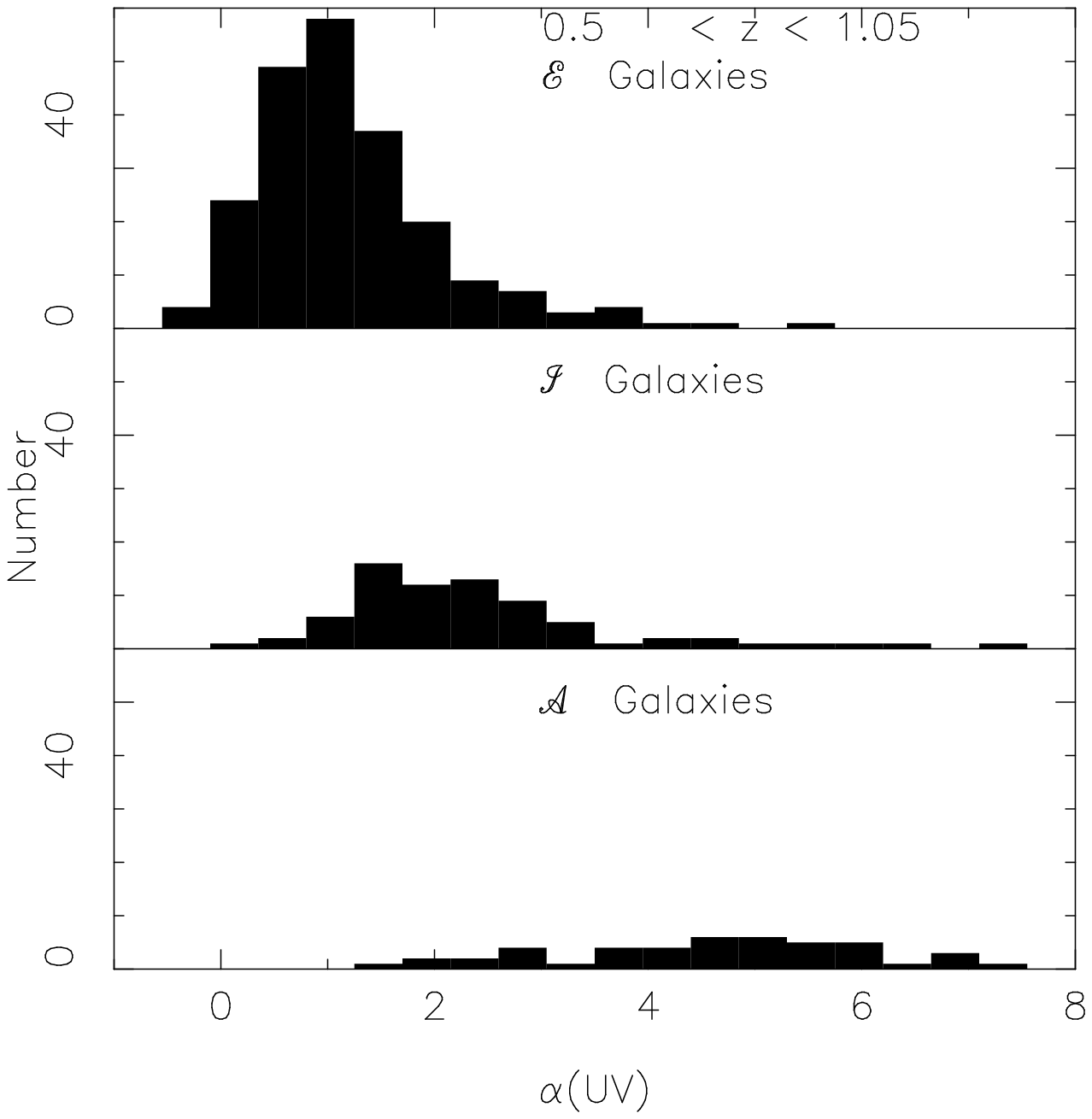}{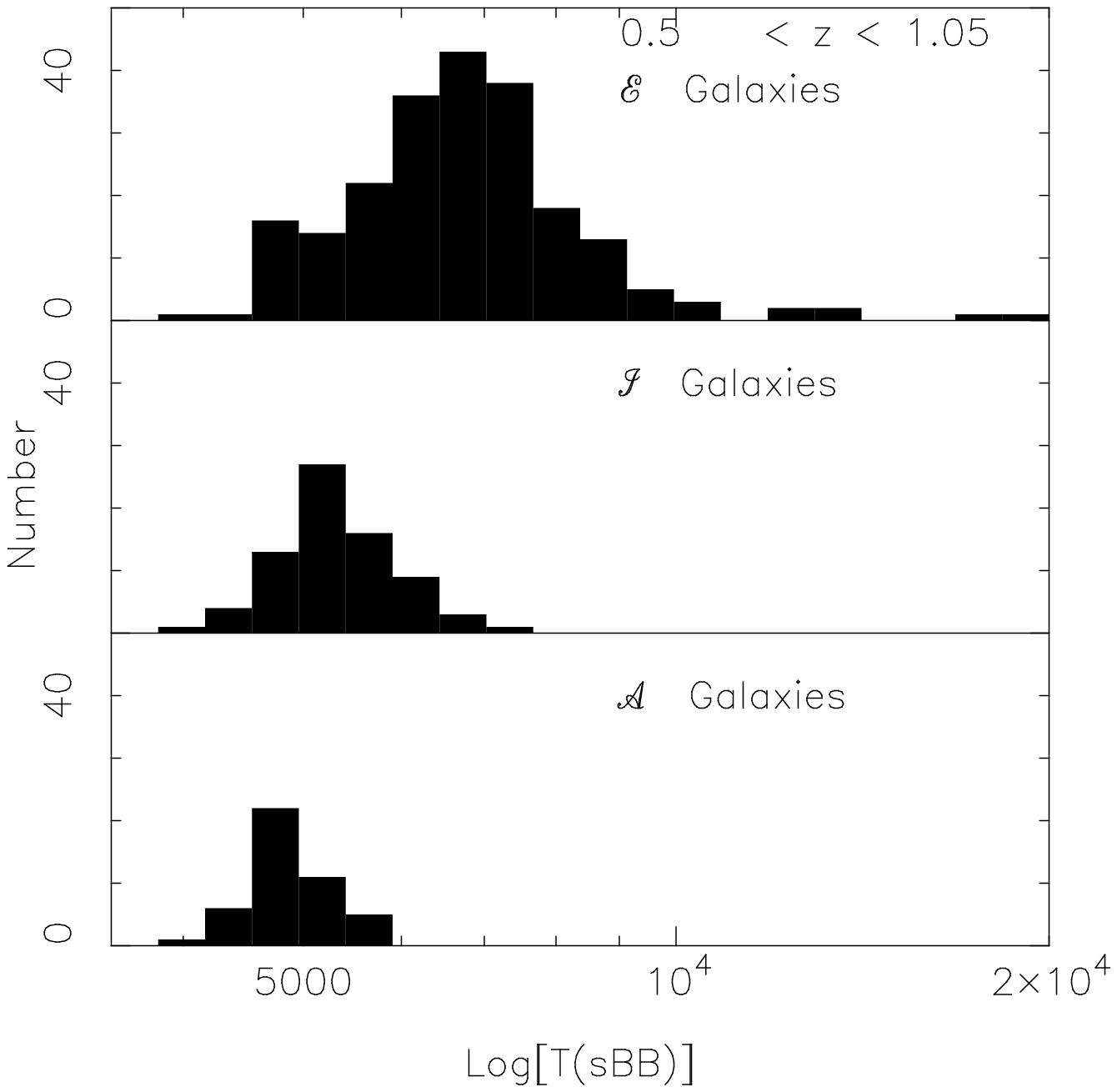}
\caption[figure8.ps]{Histograms of the the spectral indices
$\alpha_{UV}$ and $T(sBB)$ are shown
in Figures 8a and 8b 
for galaxies of spectral classes $\cal  A$, $\cal  I$ and $\cal  E$
in the ``mid'' and ``high'' redshift ranges.
Note that only the first letter of the galaxy spectral class is used.
\label{fig8}}
\end{figure}

\begin{figure}
\epsscale{0.9}
\plotone{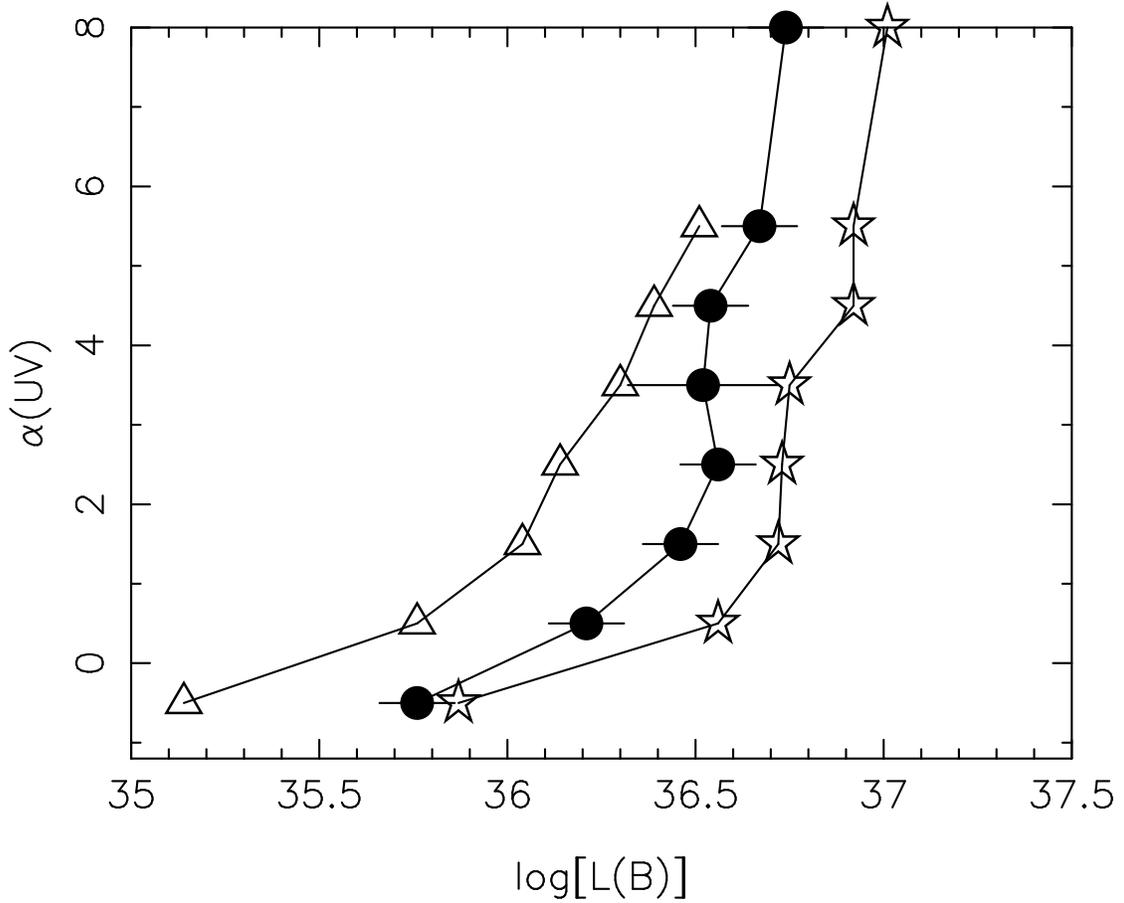}
\caption[figure9.ps]{The relationship between the median luminosity
$L(B)$ in bins of rest frame UV spectral slope $\alpha(UV)$ is shown
as a function of redshift for three of the four redshift bins.
(The highest redshift bin is not shown.)  The open triangles denote the sample
of galaxies in the region of the HDF with $0.25 < z < 0.5$, the filled
circles denote galaxies with $0.5 < z < 0.8$, and the stars denote
galaxies with $0.9 < z < 1.05$.  The range in luminosity within a given
bin in UV spectral slope is indicated by the $\pm$1$\sigma$ 
horizontal bars displayed
for the second redshift range only.
More luminous galaxies are seen to be redder
in the rest frame UV.
\label{fig9}}
\end{figure}

\begin{figure}
\epsscale{0.9}
\plotone{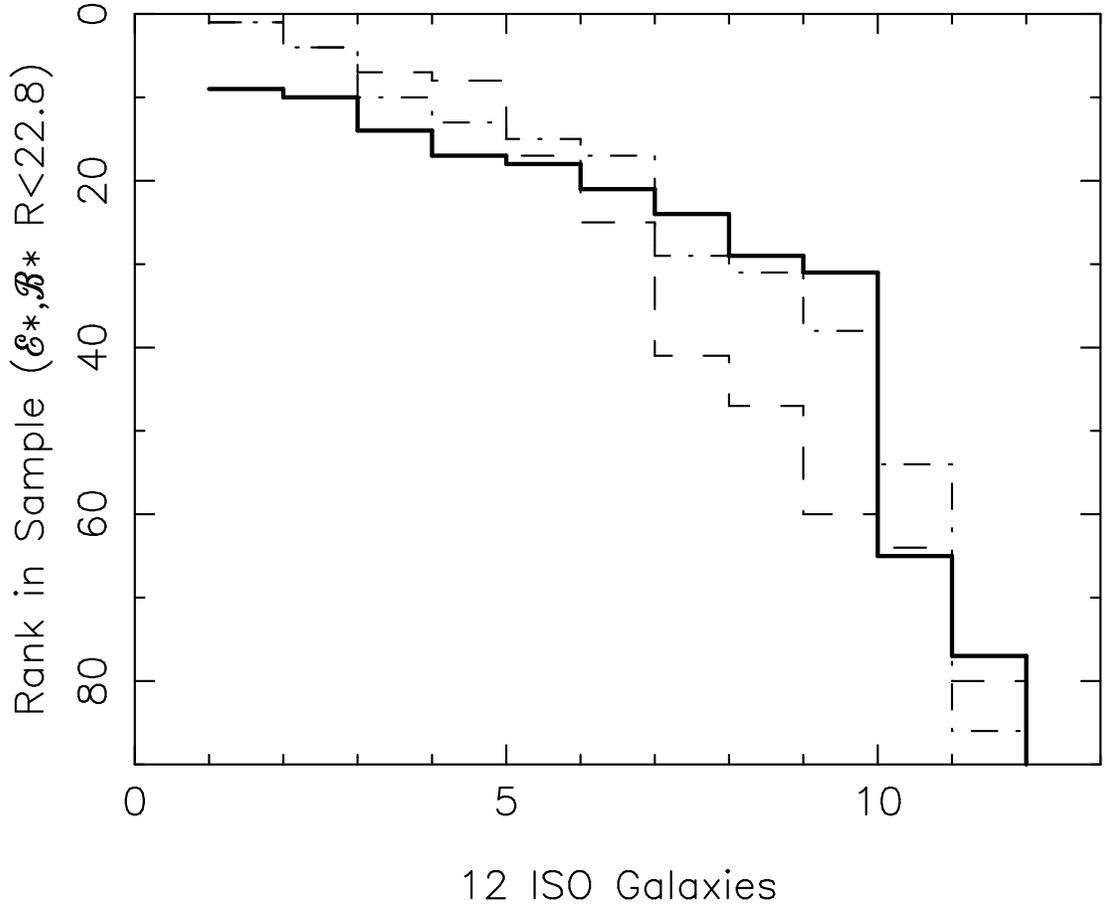}
\caption[figure10.ps]{The ranking of the galaxies detected by ISO
within the sample of all (154) galaxies with spectra showing strong
emission lines and with $R < 22.8$ is shown for the redshift 
range $0.25 < z < 1.05$.
The solid line indicates the ranking of the ISO galaxies
within the sample using $L(B)$, the dashed line is for $L(R)$,
and the dot-dashed line utilizes the mid-IR luminosity
predicted using the formalism of Meurer \etal\ (1999).  
\label{fig10}}
\end{figure}

\begin{figure}
\epsscale{1.0}
\plotone{figure11.ps}
\caption[figure11.ps]{$L(K)$ is shown as a function of the
cosmological comoving volume for galaxies
in the region of the HDF with secure redshifts.  The plot 
symbols are the same as in Figure 7.  The lines at the top
of the distribution represent the track of a galaxy of
a $L^*$ (at $K$) galaxy
with $L = 10^{11} L\subsun$ (at $z=0$) of type E, Sa and Sc with
evolutionary corrections at $K$ calculated by
Poggianti (1997) applied for $z>0$.  
\label{fig11}}
\end{figure}

\clearpage
%
%

%
%


\clearpage

%
%
\begin{deluxetable}{lrrrcrrr}
\tablenum{2}
\tablewidth{0pt}
\tablecaption{Median Spectral Indices for Objects in the Region of the HDF Using the
2p SED Model}
\label{tab2}
\tablehead{\colhead{Number} & \colhead{$\cal  A$} & \colhead{$\cal  I$} &
\colhead{$\cal  E$} & \vline & \colhead{$\cal  A$} & \colhead{$\cal  I$} &
\colhead{$\cal  E$} \nl
\colhead{} & \colhead{} & \colhead{$\alpha_{UV}$} & \colhead{} & \vline &
\colhead{} & \colhead{$\alpha_{IR}$} \nl
}
\startdata
$0.25 < z < 0.5$ \nl
 155 & 4.5 (0.6)\tablenotemark{a} & 2.6 (0.1) & 1.8 (0.1) & \vline &
1.0 (0.1) & 1.0 (0.1) & 0.6 (0.1) \nl
$0.5 < z < 0.8$ \nl
199 & 5.3 (0.2) & 2.6 (0.1) & 1.4 (0.1) & \vline &
1.4 (0.1) & 1.2 (0.1) &  0.6 (0.1) \nl
$0.8 < z < 1.05$ \nl
 135 & 5.1 (0.7) & 2.8 (0.3) & 1.3 (0.1) & \vline &
1.3 (0.1) & 1.3 (0.1) &  0.8 (0.1) \nl
$1.05 < z < 1.5$ \nl
 26 & 5.3 (1.0) & ...  & 1.1 (0.2) & \vline & 1.3 (0.2) & ...  &  1.1 (0.1) \nl
\enddata
\tablenotetext{a}{All entries have the median followed 
in parentheses by the 1$\sigma$
uncertainty of the median calculated assuming 
Gaussian statistics holds.}
\end{deluxetable}

%
%
\begin{deluxetable}{lrrrcrrr}
\tablenum{3}
\tablewidth{0pt}
\tablecaption{Median Spectral Indices for Objects in the Region of the HDF
Using the sBB SED Model}
\label{tab3}
\tablehead{\colhead{Number} & \colhead{$\cal  A$} & \colhead{$\cal  I$} &
\colhead{$\cal  E$} & \vline & \colhead{$\cal  A$} & \colhead{$\cal  I$} &
\colhead{$\cal  E$} \nl
\colhead{} & \colhead{} & \colhead{$\alpha_{UV}$} & \colhead{} & \vline &
\colhead{} & \colhead{$T$} \nl
}
\startdata
$0.25 < z < 0.5$ \nl
 155 & 4.4 (0.7)\tablenotemark{a} & 2.2 (0.1) & 1.4 (0.1) & \vline &
   5125 (100) & 5400 (100) & 6575 (100) \nl
$0.5 < z < 0.8$ \nl
199 & 4.8 (0.3) & 1.9 (0.1) & 1.0 (0.1) &  \vline &
   4875 (100) & 5300 (100) & 6600 (100) \nl
$0.8 < z < 1.05$ \nl
 135 & 4.7 (0.6) & 2.8 (0.3) & 1.0 (0.1) &  \vline &
    4900 (100) & 4950 (100) & 6700 (100) \nl
$1.05 < z < 1.5$ \nl
 26 & 4.3 (1.0) & ...  & 0.7 (0.2) &  \vline &
   4600 (300) & ...  & 6100 (200) \nl
\enddata
\tablenotetext{a}{All entries have the median followed 
in parentheses by the 1$\sigma$
uncertainty of the median calculated assuming 
Gaussian statistics holds.}
\end{deluxetable}

%
%
\begin{deluxetable}{lrrrrrclr}
\tablenum{6b}
\tablewidth{0pt}
\tablecaption{Additional Redshifts for Galaxies in the HDF}
\label{tab6b}
\tablehead{\colhead{RA} &  \colhead{Dec\tablenotemark{a}} &
\colhead{} & 
\colhead{$R$} & \colhead{$z$} & \colhead{Quality} &
\colhead{Sp.Type} & \colhead{Source}   \nl
\colhead{ ($-12$h)} & 
\colhead{($-62^{\circ}$)} & \colhead{} &
\colhead{(mag)} & \colhead{} & \colhead{}  \nl
}
\startdata
 36~~37.77 &  12 &  35.1 &  23.87 &  0.485 &   1 & $\cal  E $ &   cal \nl
 36~~45.30 &  11 &  42.9 &  24.00 &  0.558 &   1 & $\cal  E $ &   st0,cal \nl
 36~~47.17 &  13 &  41.7 &  23.93 &  1.313 &   1 & $\cal  EA$ &   
daw\tablenotemark{c}  \nl
 36~~43.84 &  12 &  41.5 &  25.21 &  4.540 &  11 & $\cal  E $ &  
stern\tablenotemark{d}  \nl
 36~~44.66 &  11 &  50.2 &  26.50 &  4.580 &  11 & $\cal  E $ & stern\tablenotemark{d} \nl
\tablenotetext{a}{J2000 coordinates}
\tablenotetext{b}{M$_B = -21.0$ (rest frame) $\equiv$ log[$L(B)$ (W)] = 36.9.}
\tablenotetext{c}{Private communication, to be published by
Dawson, Stern, Bunker, Spinrad \& Dey (2000).}
\tablenotetext{d}{Published in Stern \& Spinrad (1999).}
\enddata
\end{deluxetable}

%
%
\begin{deluxetable}{lrr rrr rrrr}
\tablenum{7b}
\tablewidth{0pt}
\tablecaption{SED Parameters for Additional Galaxies in the HDF}
\label{tab7b}
\tablehead{\colhead{RA} &  \colhead{Dec\tablenotemark{a}} &
\colhead{} &  \colhead{log[$L(B)$]} &
\colhead{$\alpha_{UV}$} & \colhead{$\alpha_{IR}$} &
\colhead{$\alpha_{UV}$} &  \colhead{log(L($\lambda_m$:{\rm blue})} &
\colhead{$T$} & \colhead{log(L($\lambda_m$:{\rm red})} \nl
\colhead{ ($-12$h)} & 
\colhead{($-62^{\circ}$)} & \colhead{} &
\colhead{(W)\tablenotemark{b}} & \colhead{(2p)} &
\colhead{(2p)} &  \colhead{(sBB)} & \colhead{(W)} & \colhead{(sBB)} &
\colhead{(W)} \nl
}
\startdata
 36~~45.30 &  11 &  42.9 &   35.67  &   1.31  &   0.59 
      &   $-$0.03 &   35.43 &  6612 &   35.57  \nl
 36~~47.17 &  13 &  41.7 &  36.69  &   0.80  &   1.14 
       &    0.68 &   36.64 &  7056 &   36.73 \nl
\tablenotetext{a}{J2000 coordinates}
\tablenotetext{b}{M$_B = -21.0$ (rest frame) $\equiv$ log[$L(B)$ (W)] = 36.9.}
\enddata
\end{deluxetable}

%
%
\begin{deluxetable}{lrrrrrcl}
\tablenum{8}
\tablewidth{0pt}
\tablecaption{Corrections to Redshifts for Galaxies in the HDF}
\label{tab8}
\tablehead{\colhead{RA} &  \colhead{Dec\tablenotemark{a}} &
\colhead{} & 
\colhead{$R$} & \colhead{Adopted $z$} & \colhead{Quality} &
\colhead{Sp.Type}   \nl
\colhead{ ($-12$h)} & 
\colhead{($-62^{\circ}$)} & \colhead{} &
\colhead{(mag)} & \colhead{} & \colhead{} & \colhead{}   \nl
}
\startdata
 36~~39.60 &  12 &  30.2 &  24.40 &  3.479 &  11 & $\cal  E $ \nl
 36~~49.43 & 13 & 16.5 & 23.63 & 1.238 & 3 & $\cal  E $ \nl
 36~~56.10 & 13 & 29.6 & 23.80 & 1.238 & 4 & $\cal  E $ \nl
\tablenotetext{a}{J2000 coordinates}
\enddata
\end{deluxetable}


\clearpage

\begin{references}

\reference{} Aaronson, M., 1978, \apjl, 221, 103

\reference{} Adelberger, K. \& Steidel, C.~C., 2000, \apj, 544, 218

\reference{} Aussel, H., Cesarsky, C.J., Elbaz, D. \& Starck, J.L., 1999, 
A\&A, 342, 313

\reference{} Barger, A.~J., Cowie, L.~L., Trentham, N., Fulton, E.,
Hu, E.~M., Songaila, A. \& Hall, D., 1998, \aj, 117, 102

\reference{} Barger, A.~J., Cowie, L.~L. \& Richards, E.~A., 2000,
\aj, 119, 2092

\reference{} Bertin, E. \& Arnouts, S., 1996, A\&AS, 117, 393

\reference{} Blain, A.~W., Kneib, J.~P., Ivison, R.~J. \& Smail, I.,
1999, \apj, 512, L87

\reference{} Blanton, M.~R. \etal, 2000, \aj\ (submitted) (Astro-ph/0012085)

\reference{} Boselli, A., Gavazzi, G., Donas, J. \& Scodeggio, M.,
2000, A\&A (in press), Astro-ph/0011016


\reference{} Brown, W.~R., Kenyon, S.~J., Geller, M.~J. \& Fabricant, D.~G.,
2000, \apjl, 540, L83

\reference{} Brunner, R.~J., Szalay, A.~S. \& Connolly, A.~J., 2000,
\apj, 541, 527

\reference {} Budav\'ari, T., Szalay, A.S., Connolly, A.J., Csabai, I. \&
Dickinson, M., 2000, \aj, 120, 1588

\reference{} Calzetti, D., 1997, \aj, 113, 162

\reference{} Calzetti, D., Kinney, A.~L. \& Storchi-Bergmann, T., 1994,
\apj, 429, 582

\reference{} Carlberg, R.~G., Cohen, J.~G., Patton, D.~R., \etal\,
2000, \apj, 532, L1

\reference{} Moriondo, G., Cimatti, A. \& Daddi, E., 2000,
A\&A, in press (Astro-ph/0010335)

\reference{} Cohen, J.~G.,  Blandford, R., Hogg, D.~W., Pahre, M.~A. \&
Shopbell, P.~L., 1999a, \apj, 512, 30

\reference{} Cohen, J.~G., Hogg, D.~W., Pahre, M.~A., Blandford, R., 
Shopbell, P.~L. \& Richberg, K., 1999b, \apjs, 120, 171


\reference{} Cohen, J.~G., Hogg, D.~W., Blandford, R., Cowie, L.~L.,
Hu, E., Songaila, A., Shopbell, P. \& Richberg, K., 2000, \apj, 538, 29

\reference{} Cole, S. \etal, 2001, \mnras, in press (Astro-ph/0012429)

\reference{} Coleman, G.~D., Wu. C.~C. \& Weedman, D.~W., 1980, \apjs, 43, 393

\reference{} Connolly, A.~J., Szalay, A.~S., Dickinson, M., 
SubbaRao, M.~U. \& Brunner, R.~J., 1997, \apj, 486, L11

\reference {} C\^ot\'e, P., Oke, J.B. \& Cohen, J.G., 1999, \aj, 118, 1645

\reference{} Cowie, L.~L., Songaila, A., Hu, E.~M. \& Cohen, J.~G., 1996,
\aj, 112, 839

\reference{} Dey, A., Graham, J.~R., Ivison, R.~J., Smail, I.,
Wright, G.~S. \& Liu, M.~C., 1999, \apj, 519, 610

\reference{} Dickinson, M.~E., 2000, in {\it Building Galaxies: From the
Primordial Universe to the Present}, XIXth Moriond Astrophysical Meeting,
eds. Hammer, F. \etal, (Paris: Ed. Frontievers), p257 (Astro-ph/0004027)

\reference{} Ellis, R.~S., Colless, M., Broadhurst, T., Heyl, J. \&
Glazebrook, K., 1996, \mnras, 280, 235

\reference{} Elmegreen, B.G., 2000, \apj, 539, 342

\reference{} Fernandez-Soto, A., Lanzetta, K.~M. \& Yahil, A., 1999, \apj, 513, 34

\reference{} Fernandez-Soto, A., Lanzetta, K.M., Chen, H.W., Pascarelle, S. \& 
Yahata, N., 2000, Astro-ph/0007447

\reference{} Folkes, S. \etal\, 1999, \mnras, 308, 459

\reference{} Frogel, J.~A., Persson, S.~E., Aaronson, M. \& Matthews, K., 1978,
\apj, 220, 75

\reference{} Fukugita, J., Shimasaku, K. \& Ichikawa, T., 1995, \pasp, 107, 945

\reference{} Gardner, J.~P., Sharples, R.~M., Frenk, C.~S., 
Carrasco, B.~E. 1997, \apj, 480, L99

\reference{} Hammer, F., Flores, H., Lilly, S.~J., Crampton, D.,
Le F\`evre, O., Rola, C., Mallen-Ornelas, G.,
Schade, D. \& Tresse, L., 1997, \apj, 481, 49

\reference{} Hogg, D. W., Cohen J. G. \& Blandford R., 2000, \apj, 545, 32

\reference{} Hogg, D. W., Cohen J. G., Blandford R., Gwyn S. D. J., 
Hartwick F. D. A., Mobasher B., Mazzei P., Sawicki M., Lin H., 
Yee H. K. C., Connolly A. J., Brunner R. J., Csabai I., 
Dickinson M., SubbaRao M. U. \& Szalay A. S., 1998, \aj, 115, 1418

\reference{} Hogg, D. W., Cohen J. G., Blandford R. \& Pahre, M.~A., 1998,
\apj, 498, L59

\reference{} Hogg, D.~W., Neugebauer, G., Cohen, J.~G., Dickinson, M., Djorgovski, S.~G.,
Matthews, K. \& Soifer, B.~T., 2000, \aj, 119, 1519

\reference{} Hogg D.~W., 2000, \aj, submitted (Astro-ph/0004054)

\reference{} Hogg D.~W., Pahre M.~A., Adelberger K.~L., Blandford R., Cohen J.~G.,
Gautier T.~N., Jarrett T., Neugebauer G. \& Steidel C.~C., 2000, 
\apjs, 127, 1 (H00)

\reference{} Hornschemeier, A.~E., \etal\, 2000, \apj, 541, 49

\reference{} Hughes, D., \etal, 1998, Nature, 394, 241


\reference{} Lilly, S.~J., Eales, S.~A., Gear, W.~K.~P., Hammer, F.,
Le F\`evre, O., Crampton, D., Bond, J.~R. \& Dunne, L., 1999, 
\apj, 518, 641


\reference{} Lin, H., Kirshner, R.~P., Shectman, S.~A., Landy, S.~D.,
Oemler, A. \& Shechter, P.~L., 1996, \apj, 464, 60

\reference{} Loveday, J., 2000, \mnras, 312, 557

\reference{} Meurer, G.~R., Heckman, T. \& Calzetti, D., 1999, \apj, 521, 64


\reference{} Oke, J.~B.,  Cohen, J.~G., Carr, M., Cromer, J., 
Dingizian, A., Harris, F.~H.,
Labrecque, S., Lucinio, R., Schaal, W., Epps, H., \& Miller, J. 1995, \pasp,
107, 307

\reference{} Persson, S.~E., Murphy, D.~C., Krzeminsky, W., Roth, M.
\& Rieke, M.~J., 1998, \aj, 116, 2475

\reference{} Poggianti, B.~M., 1997, A\&A Supl, 122, 399

\reference{} Press, W.H., Flannery, B.P., Teukolsky, S.A. \& Vetterline, W.T., 1986, Numerical Recipes, Cambridge University Press

\reference{} Richards, E.~A., 2000, \apj, 533, 611

\reference{} Richards, E.~A., Fomalont, E.~B., Kellerman, K.~I., 
Windhorst, R.~A. \& Partridge, R.~B., 1998, \aj, 116, 1039

\reference{} Richards, E.~A., Fomalont, E.~B., Kellerman, K.~I., Windhorst, R.~A.,
Partridge, R.~B., Cowie, L.~L. \& Barger, A.~J., 1999, \apj, 526 L73

\reference{} Sanders, D.~B. \& Mirabel, F., 1996, \araa, 34, 749

\reference{} Scodeggio, M. \& Silva, D., 2000, A\&A, 359, 953

\reference{} Schlegel, D.~J., Finkbeiner, D.~P. \& Davis, M., 1998, \apj, 500, 525

\reference{} Smail, I., Ivison, R.~J., Kneib, J.~P., Cowie, L.~L.,
Blain, A.~W., Barger, A.~J., Owen, F.~N. \& Morrison, G.,
1999, \mnras, 308, 1061

\reference{} Steidel, C.C. \& Hamilton, D., 1993, \aj, 105, 2017

\reference{} Steidel, C.C., Adleberger, K.~L., Giavalesco, M.,
Dickinson, M. \& Pettini, M., 1999, \apj, 519, 1

\reference{} Stern, D. \& Spinrad, H., 1999, \pasp, 111, 1475

\reference{} Stiavelli, M. \& Treu, T., 2000, to appear in
``Galaxy Disks and Disk Galaxies'', ASP Conf. Series (Astro-ph/0010100)

\reference{} Sullivan, M., Treyer, M.A., Ellis, R.S., Bridges, T.J.,
Milliard, B. \& Donas, J., 2000, 312, 442

\reference{} van den Bergh, S., Cohen, J.~G., Hogg, D.~W. \& Blandford, R.,
2000, \aj, 120, 2190

\reference{} Wainscoat, R.~J. \& Cowie, L.~L., 1992, \aj, 103, 332

\reference{} Williams, R.~E., \etal\ 1996, \aj, 112, 1335

\reference{} Worthey, G.~W., 1994, \apjs, 95, 107

\reference{} Zaritsky, D., Kennicutt, R.C.Jr \& Huchra, J.P., 1994, \apj, 
420, 87

\end{references}
\end{document}